\documentclass[12pt]{article} 
\usepackage{amsmath,amssymb,amsfonts}
\usepackage{setspace}
\usepackage{psfrag}
\usepackage{xcolor}
\usepackage{color}
\definecolor{darkblue}{rgb}{0.1,0.1,.7}
\usepackage[colorlinks, linkcolor=darkblue, citecolor=darkblue, urlcolor=darkblue, linktocpage]{hyperref} 
\usepackage[square, comma, compress,numbers]{natbib}
\usepackage[]{amsmath}
\usepackage[]{graphicx}
\usepackage[]{latexsym}
\usepackage{geometry}

\makeatletter
\g@addto@macro\@floatboxreset{\centering}
\makeatother

\usepackage{mathtools} 

\usepackage[utf8]{inputenc} 

\usepackage{booktabs,multirow}

\usepackage{amscd}
\usepackage[all,cmtip]{xy}
\usepackage{mathrsfs}
\usepackage{dsfont}

\usepackage{anyfontsize}

\usepackage{changepage}
\usepackage{mymacros}

\newcommand\preprint[1]{\gdef\@preprint{\hfill #1}}
\preprint{UUITP-04/21}

\DeclareRobustCommand{\descriptionBlackDashedKinks}[1]{\IfEq{#1}{1}{The black dashed line indicates the position of a kink.}{The black dashed lines indicate the positions of  \numberstringnum{#1} kinks.}}

\newcommand{\descriptionNColors}{The blue, orange, green, red and purple lines correspond respectively to \mbox{$N=4,5,10,20,100$}. }
\newcommand{\descriptionNColorsPlus}{The blue, orange, green, red, purple and brown lines correspond respectively to \mbox{$N=4, 5, 10, 20, 100, 10^6$}. }
\newcommand{\descriptionNColorsMinus}{The blue, orange, green, and red lines correspond respectively to \mbox{$N=4, 5, 10, 100$}. }

\newcommand{\bfu}{\mathbf{u}}
\newcommand{\bfzr}{\mathbf{0}}
\newcommand{\bfl}{\mathbf{l}}
\newcommand{\bfs}{\mathbf{s}}
\newcommand{\bfa}{\mathbf{a}}
\newcommand{\unit}{\mathds{1}}

\begin{document}

\setcounter{page}{0}
\noindent\@preprint\par
\smallskip

\vspace*{-.6in} \thispagestyle{empty}
\begin{flushright}
\end{flushright}
\vspace{.2in} {\Large
\begin{center}
{\bf Exploring $\boldsymbol{SU(N)}$ adjoint correlators in $\boldsymbol{3d}$}\\
\end{center}
}
\vspace{.2in}
\begin{center}
{\bf 
Andrea Manenti$^{a}$,
Alessandro Vichi$^{b}$} 
\\
\vspace{.2in} 
\small
$^a${\it Department of Physics and Astronomy, Uppsala University, \\Box 516, SE-751 20 Uppsala, Sweden}\\
$^b$ {\it Dipartimento di Fisica dell'Universit\`a di Pisa and INFN, \\Largo Pontecorvo 3, I-56127 Pisa, Italy}
\end{center}

\vspace{.2in}

\begin{abstract}
\noindent We use numerical bootstrap techniques to study correlation functions of scalars transforming in the adjoint representation of $SU(N)$ in three dimensions. We obtain upper bounds on operator dimensions
 for various representations and study their dependence on $N$. We discover new families of kinks, one of which could be related to bosonic QED${}_3$. 
We then specialize to the cases $N=3,4$, which have been conjectured to describe a phase transition respectively in the ferromagnetic complex projective model $CP^2$ and the antiferromagnetic complex projective model $ACP^{3}$. Lattice simulations provide strong evidence for the existence of a second order phase transition, while an effective field theory approach does not predict any fixed point. We identify a set of assumptions that constrain operator dimensions to small regions overlapping with the lattice predictions. 

\end{abstract}

\newpage

\tableofcontents

\newpage


\section{Introduction}
\label{sec:background}

The classification of 3D Conformal Field Theories (CFTs) has seen a remarkable amount of progress in the recent years thanks to the conformal bootstrap~\cite{Rattazzi:2008pe,Rychkov:2009ij} (see \cite{Poland:2018epd,Chester:2019wfx} for a review). This technique can be used to put rigorous bounds on operator dimensions, which often result in a precise determination of critical exponents~\cite{Kos:2014bka,Kos:2015mba,Kos:2016ysd,Rong:2018okz,Agmon:2019imm,Chester:2019ifh,Chester:2020iyt}. The examples studied so far include multiple scalars \cite{Li:2016wdp,Nakayama:2016jhq,Li:2017ddj,Behan:2018hfx,Kousvos:2018rhl,Kousvos:2019hgc}, spinors \cite{Iliesiu:2015qra,Iliesiu:2017nrv,Karateev:2019pvw}, conserved currents \cite{Dymarsky:2017xzb,Reehorst:2019pzi} and the energy momentum tensor~\cite{Dymarsky:2017yzx}. Furthermore, several different global symmetry groups and representations have been considered~\cite{Rattazzi:2010yc,Vichi:2011ux,Poland:2011ey,Kos:2013tga,Berkooz:2014yda,Nakayama:2014lva,Caracciolo:2014cxa,Nakayama:2014sba,Chester:2014gqa,Nakayama:2014yia,Chester:2015qca,Chester:2015lej,Chester:2016wrc,Nakayama:2016knq,Iha:2016ppj,Nakayama:2017vdd,Rong:2017cow,Chester:2017vdh,Stergiou:2018gjj,Li:2018lyb,Rong:2019qer,Reehorst:2020phk}.

Four dimensional CFTs have also been the object of similar studies. For instance, the search for the conformal window in QCD-like theories has motivated the bootstrap study of scalars in the adjoint of $SU(N_f)$~\cite{Nakayama:2016knq,Iha:2016ppj}. The bounds obtained in that case look very smooth. The plots in our setup, on the other hand, will show many interesting and suggestive features.

Our work was mainly inspired by two open problems in the condensed matter literature. Lattice models of $SU(N)$ spins have very diverse dynamics for different values of $N$. A Monte Carlo study of a model with $N=3$, defined in terms of colored loops, has produced evidence of a new fixed point~\cite{Nahum:2013qha}. Similarly, a related lattice model with $N=4$ displays evidences of a continuous phase transition~\cite{Pelissetto:2017sfd}. This last example is particularly interesting because the Monte Carlo prediction seems to contradict a naive analysis based on the Landau-Ginzburg-Wilson theory. Our goal is to shed some light on the nature of these phase transitions by performing a bootstrap study of four scalars in the adjoint of $SU(N)$. Since the setup can be easily generalized to arbitrary $N$, we also perform a systematic survey for various values of $N$, focusing mostly on a large $N$ regime. This gives us the ability to compare our results to analytic predictions available in the literature, which are often obtained as a perturbative expansion in $1/N$ around a known solution.

We find two unrelated families of kinks at large $N$. A family appearing for $N\geq 4$ when we maximize the gap on the representation $[N-2,2]$\footnote{See \eqref{eq:exchangedIrreps} and below for the definition of the $[\cdots\hspace{-.05pt}]$ notation.} and a family for $N\gtrsim20$ when we maximize the gap on the adjoint. The former approaches $1$ from below, suggesting that we are witnessing the QED$_3$ fixed points studied in~\cite{Benvenuti:2019ujm}, whereas the latter does not seem to correspond to any known theory.

In order to look more closely at the case of $N=3$ we modify the setup by adding a scalar singlet operator to the correlators considered. This gives us more parameters to tune, at the expense of a more complicated system of equations. In some instances, the study of mixed correlators, together with minimal physical assumptions, has been able to produce isolated regions in parameter space~\cite{Kos:2014bka, Kos:2015mba}. We did not obtain an island in this case, but the resulting exclusion plot is consistent with the predictions of the loop model.

In the next subsections we will briefly review known results about fixed points with $SU(N)$ global symmetry, both from a lattice and a field theory perspective. This will serve as a guide for interpreting the various features showing in the bounds that we obtain. In Section~\ref{sec:setup} we set up the bootstrap problem and write down the crossing equations. In Section~\ref{sec:StudyGenN} we present the results for general $N$ while in Sections~\ref{sec:resultsSU4} and~\ref{sec:resultsSU3} we show the results for $SU(4)$ and $SU(3)$ respectively. We leave some technical details to the appendices.

\bigskip\bigskip
\begin{em}
\noindent Note added:
\bigskip\par\noindent
While finalizing our work we became aware of~\cite{Su:2021ToAppear} which partially overlaps with our results at large $N$. We would like to thank Y.-C. He, J. Rong and N. Su for sharing their work with us and for coordinating the submission to the arXiv.

\end{em}

\subsection{Lattice models}
\label{sec:LatticeAH}
We begin with a simple lattice model, the $(A)CP^{N-1}$, which is defined as a system of spins $\bf{z_x}$ taking values in the complex projective space $CP^{N-1}$, with the index $\mathbf x$ labeling the lattice site.

Equivalently, we can describe the system by considering $\bf{z_x}$ to take values in $\mathbb C^N$, with the restriction ${\bf{\bar{z}_x}} \cdot {\bf{z_x} }=1$ and the identification $\mathbf{z_x} \sim e^{i\theta}\mathbf{z_x}$ for any $\theta\in [0,2\pi)$; the latter condition can be viewed as a $U(1)$ gauge symmetry, since one can attach a phase to each spin independently, i.e. locally. The Hamiltonian can be written as 
\begin{equation}
H_{CP^{N-1}} = J \sum_{\expval{\bf{x}, \bf{y}}} \abs{\bf{\bar{z}_x} \cdot \bf{z_y}}^2\,, \label{eq:CPNLattice}
\end{equation}
where $\expval{\bf{x}, \bf{y}}$ indicates that the sum runs over pairs of nearest neighbors. For negative $J$ the system is ferromagnetic while for positive $J$ it is antiferromagnetic. 

For $N=2$, both the ferromagnetic and the antiferromagnetic models are equivalent to the $O(3)$ Heisenberg model. 

For $N=3$ the antiferromagnetic model was shown to have the same critical exponents as the $O(8)$ model~\cite{Delfino:2015gba} while for the ferromagnetic one there is evidence of a fixed point in a new universality class with $U(3)$ global symmetry~\cite{Nahum:2013qha}. The model studied in~\cite{Nahum:2013qha} is a system of loops with $N$ colors in a three-dimensional lattice, but it can be shown to be equivalent to a ferromagnetic $CP^{N-1}$. The predicted critical exponents of the loop model read
\begin{equation}
\nu_\mathrm{loop} = 0.536(13)\,,\qquad \gamma_\mathrm{loop} =  0.97(2)\,.
\end{equation}
By comparison, the $O(8)$ exponent is given by $\nu_{O(8)} = 0.85(2)$~\cite{Delfino:2015gba}. 

For $N=4$, a Monte Carlo study has determined that the phase transition of the ferromagnetic model is of first order~\cite{Pelissetto:2019zvh}. On the other hand the antiferromagnetic model exhibits a continuous phase transition with critical exponent $\nu = 0.77(5)$~\cite{Pelissetto:2017sfd}. This results seems in contrast with a Landau-Ginzburg-Wilson (LGW) analysis which does not show evidence of stable fixed points up to six loops. We briefly review the LGW approach in the next section.

If we consider a variation of the above theory where the monopole excitations are suppressed, we obtain a completely different model, called monopole-free $CP^{N-1}$ model. Its lattice formulation has the same Hamiltonian as the compact model, but the sum is only over lattice configuration with zero monopole number. A Monte Carlo study in~\cite{Pelissetto:2020yas} determined that the model for $N=25$ undergoes a continuous phase transition with critical exponent $\nu = 0.595(15)$. 

The monopole-free $CP^{N-1}$ model is related to the non-compact $CP^{N-1}$ model in that they are characterized by the same symmetries and the suppression of monopoles, but they are distinct theories. The latter is well studied in the context of deconfined quantum criticality. In particular, the model for $N=2$ describes the phase transition between the Néel phase and the valence-bond solid (VBS) phase of two-dimensional antiferromagnets~\cite{Senthil:2004aza, Senthil_2004}. Furthermore, by non-compact $CP^{N-1}$ model one typically refers to the continuum theory. We will discuss a lattice formulation of it at the end of this section.

The $(A)CP^{N-1}$ model can also be seen as a limiting case of the Abelian Higgs model. Its Hamiltonian is defined as follows
\begin{equation}
H_{\mathrm{AH}} = - JN\sum_{\mathbf{x},\mu} 2\,\mathfrak{Re}\,\bigl(\mathbf{z_x}\cdot (\lambda_{\mathbf{x},\mu}\lsp \mathbf{z}_{\mathbf{x}+\hat{\mu}})\bigr) + H_g\,,
\end{equation}
where $H_g$ is the gauge Hamiltonian. There are two different formulations of this model: compact and non-compact. They differ in the definitions of $H_g$. The compact model has
\begin{equation}
H_g = - \kappa \sum_{\mathbf{x},\mu>\nu } 2\,\mathfrak{Re}\,\bigl(
\lambda_{\mathbf{x},\mu}\,\lambda_{\mathbf{x}+\hat\mu,\nu}\,\bar\lambda_{\mathbf{x}+\hat\nu,\mu}\,\bar\lambda_{\mathbf{x},\nu}
\bigr)\,,
\end{equation}
where the sum is over the lattice plaquettes. While in the non-compact model the fundamental degree of freedom is the gauge field $\lambda_{\mathbf{x},\mu} = e^{i A_{\mathbf{x},\mu}}$ and the Hamiltonian is defined as
\begin{equation}
H_g = \frac{\kappa}2 \sum_{\mathbf{x},\mu>\nu } \left( \Delta_{\hat\mu}A_{\mathbf{x},\nu} - \Delta_{\hat\nu}A_{\mathbf{x},\mu} \right)^2\,,
\end{equation}
where $\Delta_{\hat\mu}$ is the discretized derivative. The limit $\kappa \to 0$ Abelian Higgs model corresponds to the $(A)CP^{N-1}$ model.

The continuous model, whose Lagrangian we show in Section~\ref{sec:gaugetheories}, can be shown to have a second order phase transition for sufficiently large $N> N_c$, while at lower values of $N$ the transition is of first order. $N_c$ was estimated by field theory techniques to be $N_c = 12.2(3.9)$~\cite{Ihrig:2019kfv}. A Monte Carlo study is in agreement with this estimate and puts a more stringent upper bound: $4 < N_c < 10$~\cite{Bonati:2020jlm}.

Lastly, we would like to discuss a class of lattice models for $SU(N)$ antiferromagnets. These models exhibit the features of deconfined quantum criticality and are expected to become the non-compact $CP^{N-1}$ model in the continuum. The model is defined on a bipartite square lattice with antiferromagnetic interactions $J_1$ among nearest neighbors (distance 1) and ferromagnetic interactions $J_2$ among next-to-nearest neighbor (distance $\sqrt{2}$). On each site of the even sublattice we have spins in the fundamental of $SU(N)$ and on odd sites we have spins in the antifundamental. Let us define $P_{ij}$ and $\Pi_{ij}$ as 
\begin{equation}
P_{\mathbf{xy}} = \sum_{A=1}^{N^2-1} T_\mathbf{x}^A \cdot T_\mathbf{y}^{*A}\,,\qquad
\Pi_{\mathbf{xy}} = \sum_{A=1}^{N^2-1} T_\mathbf{x}^A \cdot T_\mathbf{y}^{A}\,.
\end{equation}
where $(T^A)^a_b$ is a Hermitian traceless matrix (generator of $SU(N)$) and $\mathbf{x},\mathbf{y}$ represent the lattice sites. $P_{\mathbf{xy}}$ acts on different sublattices while $\Pi_{\mathbf{xy}}$ acts on the same sublattice. The Hamiltonian of the model reads~\cite{Kaul:2011dqx}
\begin{equation}
H = - \frac{J_1}{N} \sum_{\langle \mathbf{x,y}\rangle} P_{\mathbf{xy}} - \frac{J_2}{N}\sum_{\langle\!\langle \mathbf{x,y}\rangle\!\rangle} \Pi_{\mathbf{xy}}\,,
\end{equation}
where $\langle \mathbf{x,y}\rangle$ and $\langle\!\langle \mathbf{x,y}\rangle\!\rangle$ denote first and second neighbors respectively. This model exhibits a deconfined quantum critical point whose critical exponents for the Néel and VBS order parameter agree with large $N$ estimates (which we will discuss in Section~\ref{sec:gaugetheories})
\begin{equation}
1 +\eta_\mathrm{V}= 0.2492N+ 0.68(4)\,,\qquad \eta_\mathrm{N}= 1 + \frac{32}{\pi^2N}-\frac{3.6(5)}{N^2}\,.
\end{equation}

\subsection{The Landau-Ginzburg-Wilson effective action}
\label{sec:LGW}

In many cases of physical interest one can understand the critical behavior of a lattice system also starting from a UV description in terms of a field theory of a scalar field with only a few renormalizable interactions.  Thanks to the properties of the RG flow, if the two UV theories belong to the same universality class, they will flow to the same fixed point in the IR. 

Physically this is equivalent to identifying the order parameter that describes the fluctuations near criticality and writing an effective Hamiltonian. The order parameter is chosen such that it vanishes in the disordered phase and is nonzero in the ordered phase. Thus, it is expected to be small near criticality and it make sense to consider only the leading terms.

If one is interested in describing the phase transition observed for $ACP^{N-1}$, the order parameter $ Q^{a}_{b}$ is a tensor of $SU(N)$ transforming in the adjoint representation and odd under an additional $\mathbb Z_2$ symmetry. The LGW Hamiltonian reads:

\begin{equation}
\label{eq:HamiltonianARPN}
\mathcal{H} = \Tr[\p_{\mu} Q \p^{\mu} Q ] + r \Tr[QQ] + \frac{u_0}4 (\Tr[QQ])^2 +\frac{v_0}{4}\Tr [QQ QQ]\,.
\end{equation}
The analysis of the $\beta$-functions for the couplings $u_0$ and $v_0$ in $\varepsilon$-expansion at one loop reveals the existence of four fixed points. Two of them are well known: the free Gaussian theory ($u_0^*=v_0^*=0$) and the $O(N')$ Wilson-Fisher fixed point ($v_0^*=0$), with \mbox{$N' = N^2-1$} the total number of scalars encoded in the matrix $Q$. In addition there are two fixed points, with both coupling nonzero and $v<0$, that merge at $N=N_c$ and turn complex for $N>N_c$. A Borel resummation of the five-loop $\varepsilon$-expansion predicts  $N_c\simeq 3.54(1)$ using the series to order $\varepsilon^3$ and $N_c \simeq 3.59(2)$ at order $\varepsilon^4$ \cite{Delfino:2015gba}.

For $N=2,3$ the additional relation $\Tr Q^4 = (\Tr Q^2)^2/2$ holds. So even for $N<N_c$ the new fixed points can be mapped respectively to the $O(3)$ and $O(8)$ model. In conclusion, the LGW analysis predicts that no fixed points exist for this model besides the WF ones. This is in tension with the lattice results discussed in the previous section.

The LGW Hamiltonian for the ferromagnetic model is similar to the one above. The only difference is the addition of a cubic term $w \Tr [ Q^3 ]$ because the order parameter does not have a $\mathbb{Z}_2$ symmetry anymore. On general grounds one would expect no continuous phase transitions except the WF ones in this case as well. However, this is again in disagreement with the lattice results.

\subsection{Scalar gauge theories}
\label{sec:gaugetheories}

In this section we will briefly review the 3$d$ gauge theories relevant for our bootstrap setup. First let us focus on Abelian models. Let $\phi^a$ be a scalar transforming in the fundamental of $SU(N)$ and minimally coupled to a $U(1)$ gauge field, with the following Lagrangian
\begin{equation}
\mathcal{L}_{\mathrm{AH}} = \frac1{4e^2} F_{\mu\nu}F^{\mu\nu} + \sum_{a=1}^N\left|(\partial_\mu + i A_\mu)\phi^a\right|^2 + m^2 \sum_{a=1}^N \phi^a\phi_a + \lambda\lsp \biggl(\,\sum_{a=1}^N\phi^a \phi^*_a\biggr)^2\,.
\label{eq:AHlagrangian}
\end{equation}
This model has a fixed point for $\lambda = m^2 = 0$ which is the (tricritical) bosonic QED and a fixed point where $\lambda\neq0$ which is the Abelian Higgs model. In the limit where $e\to0$ --- namely when the gauge field becomes non dynamical --- the latter is referred to as the $CP^{N-1}$ model. The name stems from its interpretation as a nonlinear sigma model with target space $CP^{N-1}$. Indeed the quartic potential gives a constraint $\phi^a\phi_a = \mathrm{const.}$ and the non dynamical gauge field imposes the identification $\phi^a \sim e^{i\alpha} \phi^a$.

In~\cite{Benvenuti:2019ujm} we can find the values for various anomalous dimensions in both fixed points as an expansion in $1/N$. In the bosonic QED fixed point, the bilinear of $\phi$ in the adjoint and in the singlet representations read, respectively
\begin{equation}
\begin{aligned}
\Delta[|\phi|_\mathrm{adj}^2] &= 1- \frac{64}{3\pi^2 N} + O\Bigl(\frac1{N^2}\Bigr)\,,\\
\Delta[|\phi|_\mathrm{sing}^2] &= 1+ \frac{128}{3\pi^2 N} + O\Bigl(\frac1{N^2}\Bigr)\,.
\end{aligned}
\end{equation}
Similar results are available for the Abelian Higgs model. However, in place of the singlet bilinear we have the Hubbard-Stratonovich field $\sigma$, whose dimension starts with~2
\begin{equation}
\begin{aligned}
\Delta[|\phi|_\mathrm{adj}^2] &= 1- \frac{48}{3\pi^2 N} + O\Bigl(\frac1{N^2}\Bigr)\,,\\
\Delta[\sigma] &= 2 - \frac{48}{\pi^2 N} + O\Bigl(\frac1{N^2}\Bigr)\,.
\end{aligned}
\end{equation}
In the condensed matter literature $|\phi|_\mathrm{adj}^2$ is known as the Néel order parameter and the quantities that are typically quoted are the critical exponents, which are given as $\eta_{\mathrm{N}} = 2 \Delta[|\phi|_\mathrm{adj}^2] -1$ and $\nu^{-1}_{\mathrm{N}} = 3- \Delta[\sigma]$. Also note that the above formulas appeared in~\cite{Kaul:2008xw, Halperin:1973jh, Hikami:1979ih, Vasiliev:1984gn} first.

In~\cite{Benvenuti:2019ujm} it was also computed the anomalous dimension of scalar quartic operators in the representation $[N-1,N-1,1,1]$\footnote{They denote this representation $[2,0,\ldots ,0,2]$. See Subsection~\ref{sec:OPE} for the definition of our notation.}:
\begin{equation}
\begin{aligned}
\Delta[|\phi|_{[N-1,N-1,1,1]}^4] &= 2 - \frac{128}{3\pi^2 N} + O\Bigl(\frac1{N^2}\Bigr)\,, \quad \text{for bosonic QED}\,,\\
\Delta[|\phi|_{[N-1,N-1,1,1]}^4] &= 2- \frac{48}{3\pi^2 N} + O\Bigl(\frac1{N^2}\Bigr)\,, \quad \text{for Abelian Higgs}\,.
\end{aligned}\label{eq:biadjoint}
\end{equation}

There are also results for the anomalous dimensions of the monopole operators in QED-like theories~\cite{Murthy:1989ps, Metlitski:2008dw, Dyer:2015zha}; contrarily to fermionic QED${}_3$, however, monopoles operators are neutral under the global $SU(N)$ symmetry and will not appear in our setup.

We decided to focus on the region in parameter space where $\Delta \lesssim 1$, therefore fermionic gauge theories are typically not interesting for us as they contain heavier operators, namely $\Delta[\bar\psi\psi] \sim 2$. However, the model QED-GN$_+$ considered in~\cite{Benvenuti:2019ujm} contains a Hubbard-Stratonovich field which is in the same range of dimensions that we study. The Lagrangian reads
\begin{equation}
\mathcal{L}_{\mbox{\scriptsize QED-GN}} =  \frac1{4e^2} F_{\mu\nu}F^{\mu\nu} + \sum_{a=1}^N\bar{\psi}_a\,\gamma^\mu(\partial_\mu + i A_\mu)\lsp\psi^a + \rho \sum_{a=1}^N \bar{\psi}_a \psi^a + \mu^2 \rho^2\,.
\end{equation}
Integrating out $\rho$ leaves us with a four-fermion interaction. The anomalous dimension for the lightest scalar singlet is
\begin{equation}
\Delta[\rho] = 1- \frac{144}{3\pi^2 N}+ O\Bigl(\frac1{N^2}\Bigr)\,,
\end{equation}
while the adjoint this time is heavier
\begin{equation}
\Delta[(\bar\psi\psi)_\mathrm{adj}] = 2- \frac{48}{3\pi^2 N}+ O\Bigl(\frac1{N^2}\Bigr)\,.
\end{equation}

As far as we are aware, there are no such results in the context of nonabelian gauge theories. The large $N$ dynamics of QCD$_3$ is well known and studied~\cite{Armoni:2017jkl, Armoni:2019lgb, Argurio:2019tvw}, but the anomalous dimensions of low-lying operators have not been computed so far. We can however argue, by analogy with a similar class of models studied in~\cite{Sachdev:2018nbk} and previously in~\cite{Hikami:1980qk, Vasiliev:1984gn, Vasiliev:1984bf}, that for an $SU(M)$ or $SO(M)$ gauge theory with global symmetry $SU(N)$, the anomalous dimensions at large $N$ will be corrected by terms linear in $M$.

As it was first observed in \cite{Reehorst:2020phk} and then refined in \cite{Su:2021ToAppear}, the abelian or non-abelian nature of the underlining gauge theory reflects in gaps in the spectrum of scalar operators in certain representations.  
Consider for instance $\mathcal O^{[a\, b]}_{[c\, d]}$, the smallest scalar operator with two antisymmetric fundamental and two antisymmetric antifundamental indexes. We will refer to this representation as $[N-2,2]$ in the rest of the paper. Such representation appears in the OPE of two adjoint operators. In a Generalized Free Theory (GFT) of an adjoint operator $Q^a_{c}$ or in the LGW theory defined in the previous section, the smallest scalar in the $[N-2,2]$ representation is given by
\begin{equation}
\mathcal O^{[a\, b]}_{[c\, d]} \sim Q^a_{c} Q^{b}_d - Q^a_{d} Q^{b}_c\,.
\label{eq:ONm22}
\end{equation}
Similarly, given  a non-abelian gauge theory with fundamental fields $\phi^A_a$ one can define  the gauge invariant scalar $Q^a_{c} \sim \phi^{\dagger A a} \phi^{ A}_c - \text{trace}$. Then, the combination in \eqref{eq:ONm22}  is non-trivial due to the internal gauge indexes. However, if these were absent, one could not construct it. In the Abelian Higgs model \eqref{eq:AHlagrangian} the smallest non trivial operator in the $[N-2,2]$ representation that one can construct requires two extra derivatives or more fields.
This reasoning is valid only in a neighbourhood of the UV description, however it gives us an intuition about which operators we should expect in the CFT. Hence, we do not expect the IR fixed point of abelian gauge theories or $CP^N$ models to have light scalars in the $[N-2,2]$ representation.


\section{Setup}

\label{sec:setup}

\newcommand{\listIrreps}{\{S,T^2,T^4,A^2,H,B\}}

In this section we derive the crossing equations for the four-point function of a scalar operator $Q_a^b$ transforming in the adjoint of $SU(N)$. In order to keep the notation compact we use an index-free formalism for the four-point structures. This means that we contract any fundamental index $a$ with a complex polarization $\bar{S}^{a}$ and and any anti-fundamental index $b$ with a complex polarization $S_b$. Then the operator $Q$ becomes
\be
Q(x,S,\bar{S}) = \bar{S}^{a} S_{b} Q_a^b\,.
\ee
The tracelessness properties of the operator can be encoded in the requirement $S\cdot\bar{S}=0$. In order to discuss the OPE it is easier to use a formalism with explicit indices instead. In the following we will use $\mathbf{p}_i$ as a shorthand for the point $x_i$ together with the polarizations that are attached to it.

\subsection[\texorpdfstring{The $Q\times Q$ OPE}{The Q x Q OPE}]{The $\boldsymbol{Q\times Q}$ OPE}
\label{sec:OPE}

We consider the single correlator of four $Q$'s. The representations exchanged in the OPE $Q \times Q$ were studied in~\cite{Iha:2016ppj}. Here we rederive their results in our notation. The exchanged representations are
\begin{equation}
\begin{aligned}
[N-1,1] \otimes [N-1,1] &=\bullet_+  \oplus [N-2,1,1]_- 
 \oplus [N-1,N-1,2]_- \oplus [N-2,2]_+ \\& \oplus [N-1,N-1,1,1]_+  \oplus\, [N-1,1]_+ \oplus [N-1,1]_-\,.
\end{aligned} \label{eq:exchangedIrreps}
\end{equation}
We adopt the notation $[\lambda_1,\lambda_2,\cdots]$ to denote a Young tableau with $\lambda_1$ boxes in the first column, $\lambda_2$ boxes in the second column and so on. For brevity we also denote the singlet (i.e. the empty tableau) as $\bullet$. The subscript $\pm$ refers to the symmetry property under exchanging the two operators and, therefore, to the parity of the spin of the operators entering the OPE. Also note that the representations $[N-2,1,1]$ and $[N-1,N-1,2]$ are conjugate to each other.

We first write down the two-point tensor structures of all exchanged operators and then the tensor structures entering the OPE. By taking an OPE inside the $\langle QQQQ\rangle$ correlator we can construct the four-point structures.

\subsection{Two-point functions}

Here we will fix the convention for the two-point functions of all the operators exchanged. We also denote as $\mathcal{T}_2$ the unique two-point structure of two real symmetric traceless operators with dimension $\Delta$ and spin~$\ell$
\begin{equation}
\mathcal{T}_2 = \frac{I^{\mu_1}_{(\nu_1}\cdots I^{\mu_\ell}_{\nu_\ell)} - \mathrm{traces}}{(x_{12}^2)^{\Delta}}\,,\qquad I^\mu_\nu = \delta^\mu_\nu - 2 \frac{x_{12\lsp \nu} x_{12}^\mu}{x_{12}^2}\,.
\end{equation}
Letting $\mathcal{O}_{\mathbf{irrep}}$ be an exchanged operator in the representation $\mathbf{irrep}$ we have
\begin{subequations}
\begin{align}
&\langle \mathcal O_{\bullet}\mathcal{O}_{\bullet}  \rangle =  \mathcal T_2\,,\\
&\langle \mathcal O_{[N-1,1]}{}^a_b \mathcal{O}_{[N-1,1]}{}^c_d  \rangle = \mathcal T_2\,\biggl(\delta^a_d\delta^c_b - \frac{\delta^a_b\delta^c_d}{N}\biggr) \,,\\
&\begin{aligned}
\langle \mathcal O_{[N-1,N-1,1,1]}{}^{ab}_{cd} \mathcal{O}_{[N-1,N-1,1,1]}{}^{ef}_{gh}  \rangle &= \frac{\mathcal T_2}4\biggl(
\delta _{(g}^a \delta _{h)}^b \delta _{(c}^e \delta _{d)}^f
-\frac{\delta^{(a}_{(\!(h}\delta^{b)}_{(c}\delta^{(e}_{d)}\delta^{f)}_{g)\!)}}{N+2} 
+\frac{2\lsp\delta _{(c}^a \delta _{d)}^b \delta _{(g}^e \delta _{h)}^f}{(N+1)(N+2)}
\biggr) \,,
\end{aligned}\\
&\begin{aligned}
\langle \mathcal O_{[N-2,2]}{}^{ab}_{cd} \mathcal{O}_{[N-2,2]}{}^{ef}_{gh}  \rangle &= \frac{\mathcal T_2}4\biggl(
\delta _{[g}^a \delta _{h]}^b \delta _{[c}^e \delta _{d]}^f
-\frac{\delta^{[a}_{[\!\lsp[h}\delta^{b]}_{[c}\delta^{[e}_{d]}\delta^{f]}_{g]\!\lsp]}}{N-2}
+\frac{2\lsp\delta _{[c}^a \delta _{d]}^b \delta _{[g}^e \delta _{h]}^f}{(N-1)(N-2)}
\biggr) \,,
\end{aligned}\\
&\begin{aligned}
\langle \mathcal O_{[N-2,1,1]}{}^{ab}_{cd} \mathcal{O}_{[N-1,N-1,2]}{}^{ef}_{gh}  \rangle &= \frac{\mathcal T_2}4\biggl(
\delta _{[g}^a \delta _{h]}^b \delta _{(c}^e \delta _{d)}^f+\frac{\delta^{[a}_{[\!\lsp[h}\delta^{b]}_{(c}\delta^{(e}_{d)}\delta^{f)}_{g]\lsp\!]}}{N} 
\biggr)\,.
\end{aligned}
\end{align}
\end{subequations}
In the above equations we use pairs of parentheses $(\!($ or $[\!\lsp[$ to avoid ambiguities when pairs of indices being (anti)symmetrized overlap. The (anti)symmetrizations are defined as $X_{[ab]} = X_{ab}-X_{ba}$ and $X_{(ab)} = X_{ab}+X_{ba}$.

\subsection{OPE structures}

The operators in the adjoint and the singlet appear in the OPE multiplied by some Kronecker $\delta$'s in order to achieve the correct number of indices with the right symmetry and traceleness properties. All in all the OPE reads
\begin{equation}
\begin{aligned}
Q^a_c \times Q^b_d &= \sum_{\mathcal{O}}\delta^b_c\lsp \mathcal{O}^+_{[N-1,1]}{}^a_d +\delta^a_d\lsp \mathcal{O}^+_{[N-1,1]}{}^b_c - \frac2{N}\bigl(\delta^a_c\lsp \mathcal{O}^+_{[N-1,1]}{}^b_d + \delta^b_d\lsp \mathcal{O}^+_{[N-1,1]}{}^a_c \bigr) 
\\&+
\sum_{\mathcal{O}}\delta^b_c\lsp \mathcal{O}^-_{[N-1,1]}{}^a_d - \delta^a_d\lsp \mathcal{O}^-_{[N-1,1]}{}^b_c
+ \sum_{\mathcal{O}}\mathcal{O}_\bullet \lsp \Bigl(\delta^a_d\delta^b_c - \frac{\delta^a_c\delta^b_d}{N}\Bigr)
\\&+\sum_{\mbox{\scriptsize other }\mathbf{irreps}} \sum_{\mathcal{O}}\mathcal{O}_{\mathbf{irrep}}{\,}^{ab}_{cd}\,.
\end{aligned}
\end{equation}

\subsection{Four-point structures and crossing}
\label{ssec:fpf}

By taking the two-point functions of pairs of operators appearing in the OPE $Q\times Q$ we can construct the four-point structures. We contract all the indices by the polarizations $S_a$ and $\bar S^a$ as discussed at the beginning. We also denote as $\mathcal{T}_4$ the unique four-point structure of a single scalar of dimension $\Delta_Q$ 
\begin{equation}
\mathcal{T}_4 = \frac{1}{(x_{12}^2 x_{34}^2)^{\Delta_Q}}\,.
\end{equation}
The four-point function can be written in terms of six real functions of the cross ratios $u = \frac{x_{12}^2x_{34}^2}{x_{13}^2x_{24}^2}$ and $v = u|_{1\leftrightarrow3}$.
\be
\langle Q(\mathbf{p}_1) Q(\mathbf{p}_2) Q(\mathbf{p}_3) Q(\mathbf{p}_4)\rangle = \cT_4 \sum_{r=1}^6 T^{(r)}\,f_r(u,v)\,.
\ee
The structures $T^{(r)}$ are chosen so that they are associated to the exchange of a single representation, as summarized in Table~\ref{tab:frOPEsinglecorr}. Let us introduce the shorthand
\be
\mathcal{S}_{ijkl} = (S_1\cdot \bar{S}_i)(S_2\cdot \bar{S}_j)(S_3\cdot \bar{S}_k)(S_4\cdot \bar{S}_l)\,.
\ee
With this notation in mind we have
\be
\begin{aligned}
T^{(1)} &= \frac{\mathcal{S}_{(43)(21)}}4 - \frac{\mathcal{S}_{2341}+\mathcal{S}_{2413}+\mathcal{S}_{4123}+\mathcal{S}_{3142}}{4(N+2)}+\frac{\mathcal{S}_{2143}}{2(N+1)(N+2)}\,,\\
T^{(2)} &= \frac{\mathcal{S}_{4321}-\mathcal{S}_{3412}}{2} - \frac{\mathcal{S}_{2341}-\mathcal{S}_{2413}+\mathcal{S}_{4123}-\mathcal{S}_{3142}}{2N}\,,\\
T^{(3)} &=\frac{\mathcal{S}_{[43][21]}}{4}  - \frac{\mathcal{S}_{2341}+\mathcal{S}_{2413}+\mathcal{S}_{4123}+\mathcal{S}_{3142}}{4(N-2)}+\frac{\mathcal{S}_{2143}}{2(N-1)(N-2)}\,\\
T^{(4)} &=\mathcal{S}_{2341}+\mathcal{S}_{2413}+\mathcal{S}_{4123}+\mathcal{S}_{3142} - \frac{4\,\mathcal{S}_{2143}}{N}\,,\\
T^{(5)} &= -\mathcal{S}_{2341}+\mathcal{S}_{2413}-\mathcal{S}_{4123}+\mathcal{S}_{3142}\,,\\
T^{(6)} &= \mathcal{S}_{2143}\,.
\end{aligned}\label{eq:structsSingle}
\ee
Now we can write down the conformal block expansions for each of the $f^{(r)}$. For $r=1,3,4,6$ one has
\be
\begin{aligned}
f_r(u,v) &= \sum_{\substack{\cO_{\Delta,\ell} \in\, \mathbf{r}\\\ell\;\text{even}}}\,\big(\lambda_{QQ\cO}^{\mathbf{r}} \big)^2 \,g_{\Delta,\ell}(u,v)\,,\\
\end{aligned}
\ee
with $\mathbf{1}=[N-1,N-1,1,1]$, $\mathbf{3}=[N-2,2]$, $\mathbf{4}=[N-1,1]_+$ and $\mathbf{6}=\bullet$. The other representations follow:
\be
\begin{aligned}
f_2(u,v) &= \sum_{\substack{\cO_{\Delta,\ell} \in [N-2,1,1]\\\ell\;\text{odd}}}\,\lambda_{QQ\cO}^{[N-2,1,1]} \lambda_{QQ\cO}^{[N-1,N-1,2]} \,g_{\Delta,\ell}(u,v)\,,\\
f_5(u,v) &= \sum_{\substack{\cO_{\Delta,\ell} \in [N-1,1]\\\ell\;\text{odd}}}\,\big(\lambda_{QQ\cO}^{[N-1,1]_-}\big)^2\,g_{\Delta,\ell}(u,v)\,.\end{aligned}
\ee
Positivity holds for all sums because $\lambda_{QQ\cO}^{[N-2,1,1]}=\bigl(\lambda_{QQ\cO}^{[N-1,N-1,2]}\bigr)^*$. The conformal blocks are normalized according to this convention
\begin{equation}
g_{\Delta,\ell}(u,v) \underset{u\to0,v\to1}{\sim}  \left(\frac{u}{16}\right)^{\Delta/2}\,P_\ell\left(\frac{1-v}{2\sqrt{u}}\right)\,,\label{eq:cbconvention}
\end{equation}
$P_\ell$ being a Legendre polynomial. This normalization implies $g_{\Delta,\ell}\big(\frac14,\frac14\big)\geq0$ for all $\Delta,\ell$.

\begin{table}
\centering
\begin{tabular}{lllll}
\toprule
$f_r$ & $\cO$ exchanged & spin parity\\
\midrule
$f_1$ & $[N-1,N-1,1,1]$  & even \\
$f_2$ & $[N-2,1,1] + \mathrm{h.c.}$ & odd\\
$f_3$ & $[N-2,2]$  & even\\
$f_4$ & $[N-1,1]_+$ & even\\
$f_5$ & $[N-1,1]_-$ & odd\\
$f_6$ & $\bullet$ & even\\
\bottomrule
\end{tabular}
\caption{Operator contributions to the partial wave $f_r$.}\label{tab:frOPEsinglecorr}
\end{table}

We are now ready to write the crossing equations. First let us introduce the even and odd combinations of the conformal blocks
\begin{equation}
\begin{aligned}
F_{\Delta,\ell} = v^{\Delta_Q}\lsp g_{\Delta,\ell}(u,v) - u^{\Delta_Q} \lsp g_{\Delta,\ell}(v,u) \,,\\
H_{\Delta,\ell} = v^{\Delta_Q}\lsp g_{\Delta,\ell}(u,v) + u^{\Delta_Q} \lsp g_{\Delta,\ell}(v,u) \,.
\end{aligned}
\end{equation}
The crossing equations then read
\begin{equation}
\begin{aligned}
V_{\bullet;\lsp0,0} &= \sum_{r=1,3,4,6}  \sum_{\substack{\cO_{\Delta,\ell} \in\, \mathbf{r}\\\ell\;\text{even}}} \bigl(\lambda^{\mathbf r}_{QQ\cO}\bigr)^2\lsp V_{\mathbf{r};\lsp\Delta,\ell} + \sum_{\substack{\cO_{\Delta,\ell} \in [N-2,1,1]\\\ell\;\text{odd}}} \,\bigl|\lambda_{QQ\cO}^{[N-2,1,1]} \bigr|^2\, V_{[N-2,1,1];\lsp\Delta,\ell} \\ &+
 \sum_{\substack{\cO_{\Delta,\ell} \in [N-1,1]\\\ell\;\text{odd}}}\,\big(\lambda_{QQ\cO}^{[N-1,1]_-}\big)^2\, V_{[N-1,1]_-;\lsp\Delta,\ell}\,,
\end{aligned}
\end{equation}
where we defined the vectors $V_{\mathbf{irrep};\lsp\Delta,\ell}$ as
\renewcommand{\arraystretch}{1.2}
\begin{equation}
\begin{aligned}
&V_{[N-1,N-1,1,1];\lsp\Delta,\ell} = \left(
	\begin{array}{c}
	F_{\Delta,\ell} \\ 0_3 \\ H_{\Delta,\ell} \\ 0
	\end{array}
\right)\,,
\qquad
V_{[N-2,1,1];\lsp\Delta,\ell} = \left(
	\begin{array}{c}
	0 \\ F_{\Delta,\ell} \\ 0_3 \\ H_{\Delta,\ell}
	\end{array}
\right)\,, \\&
V_{[N-2,2];\lsp\Delta,\ell} = \left(
	\begin{array}{c}
	0_2 \\ F_{\Delta,\ell} \\ 0 \\ - \frac{4(N-3)(N+1)}{(N-1)(N+3)} H_{\Delta,\ell} \\ \frac{2(N-3)N^2}{(N-2)(N-1)(N+2)} H_{\Delta,\ell}
	\end{array}
\right)\,,
\;
V_{[N-1,1]_+;\lsp\Delta,\ell} = \left(
	\begin{array}{c}
	0_3 \\ F_{\Delta,\ell} \\ -\frac{4(N-2)(N+1)(N+2)}{N^2(N+3)}H_{\Delta,\ell} \\ \frac{N+2}{N}H_{\Delta,\ell}
	\end{array}
\right)\,,
\\&
V_{[N-1,1]_-;\lsp\Delta,\ell} = \left(
	\begin{array}{c}
	-\frac{4(N+1)}{N+2}F_{\Delta,\ell} \\ \frac{2N}{(N-2)(N+2)} F_{\Delta,\ell} \\ \frac{N-1}{N-2} F_{\Delta,\ell} \\
	\frac{N^4}{(N-2)^2(N+2)^2} F_{\Delta,\ell} \\ \frac{4(N+1)}{N+3} H_{\Delta,\ell} \\ -\frac{N}{N+2} H_{\Delta,\ell}
	\end{array}
\right)\,,
\quad
V_{\bullet;\lsp\Delta,\ell} = \left(
	\begin{array}{c}
	\frac{(N-1)(N+1)}{N(N+2)}F_{\Delta,\ell} \\ \frac{(N-1)(N+1)}{2(N-2)(N+2)} F_{\Delta,\ell} \\ \frac{(N-1)(N-1)}{4 N (N-2)} F_{\Delta,\ell} \\
	\frac{N(N-1)(N+1)}{(N-2)^2(N+2)^2} F_{\Delta,\ell} \\ -\frac{4(N+1)}{N(N+3)} H_{\Delta,\ell} \\ -\frac{N+1}{N+2} H_{\Delta,\ell}
	\end{array}
\right)\,.
\end{aligned} \label{eq:CrossingEquationsSing}
\end{equation}
\renewcommand{\arraystretch}{1}

\noindent
With the notation $0_n$ we mean a sequence of $n$ zeros. If $N = 3$ we need to discard the third line and the representation $[N-2,2]$ is not exchanged. If $N = 2$ we need to discard the second, third and fourth line and the representations $[N-2,2]$ and $[N-2,1,1]$ are not exchanged.


\begin{figure}[t]
	\centering
	\includegraphics[width=0.8\textwidth]{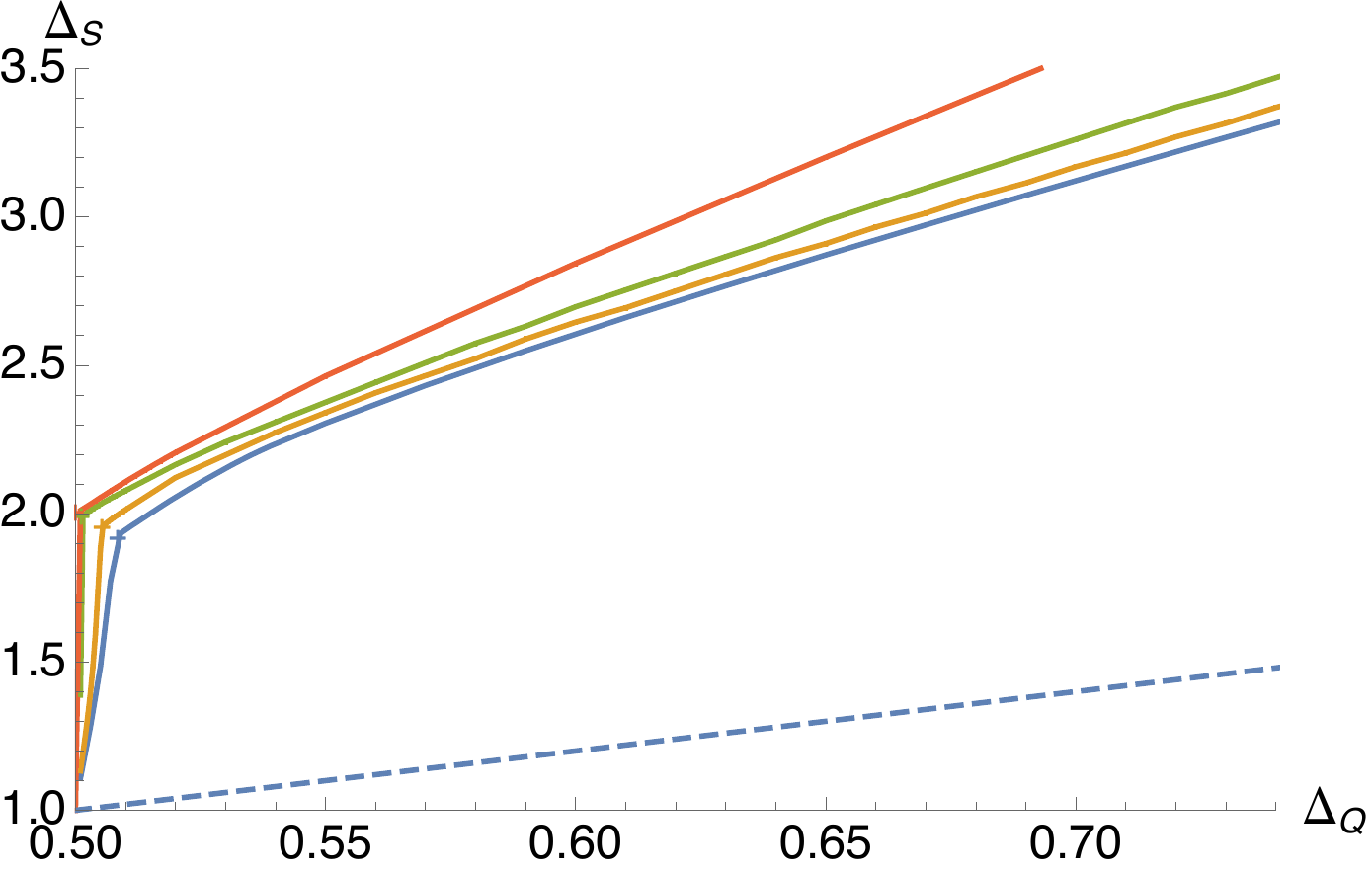}
	\caption{Bound on the dimension of the first singlet scalar.  These bounds have been obtained at $\Lambda=19$.  \descriptionNColorsMinus  The crosses corresponds to the large-$N$ prediction for the $O(N^2-1)$ model.}
	\label{fig:bisectionallnoverviewsinglet}
\end{figure}

\section{A systematic study for general \texorpdfstring{$\boldsymbol{N}$}{N}}

\label{sec:StudyGenN}
Here we present a systematic study  for general $N$ of bounds on the dimension of the first scalar operator in all representations containing even spins. Specifically we examine $N=4,5,10,100$ and occasionally other values. The bounds on the leading operators in the singlet representation are identical  to the corresponding bounds found in the $O(N')$-vector bootstrap,\footnote{This is proven in appendix \ref{app:RelationshipToVectorBootstrap}.} where $N'=N^2-1$. For what concerns other representations, any solution with $O(N')$ symmetry is also a solution of our crossing equations but the bounds do not coincide as long as we extremize a single operator at the time.

\subsection{Bounds on operator dimensions}

\subsubsection*{Singlets}

The bound on the dimension of the first singlet scalar $\Delta_S$ shows a clear kink corresponding to the $O(N')$ model under the identification $\phi^{i}\to Q^{a}_{b}$. In addition there is a second set of (very mild) kinks for $N=4,5$ which get closer as $N$ increases.  These bounds are shown in Figure \ref{fig:bisectionallnoverviewsinglet}. In the scalar singlet sector we do not find any new interesting feature.

\begin{figure}[thbp]
\includegraphics[width=0.48\textwidth]{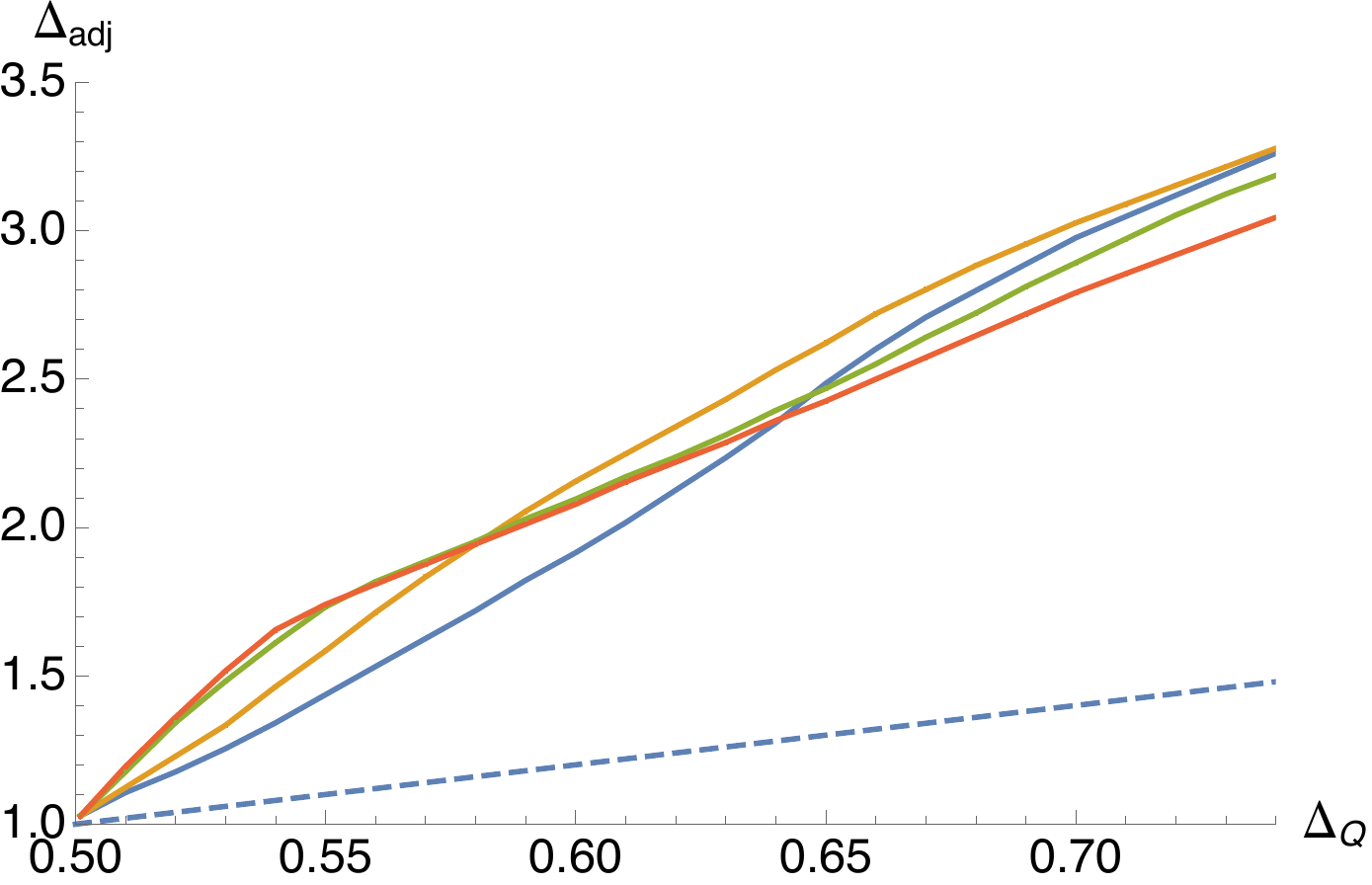}
\includegraphics[width=0.48\textwidth]{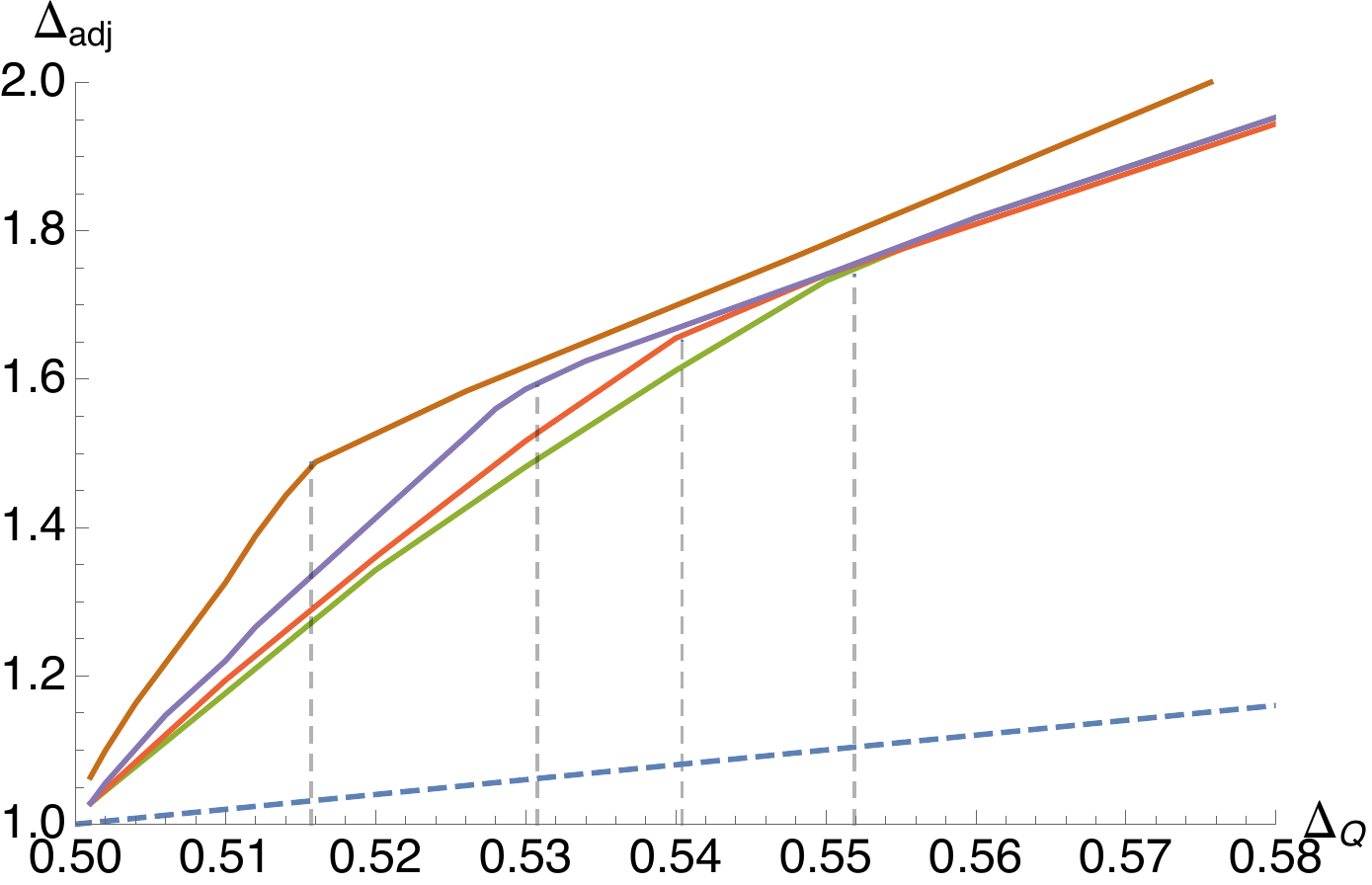}
		\caption{Bound on the dimension of the first adjoint scalar. We do not assume that  $Q$ is odd under a global $\mathbb Z_2$ symmetry. \descriptionNColorsPlus A kink appears for $N\gtrsim 20$. The bounds have been computed at $\Lambda=19$. The dashed line corresponds to Generalized Free Theory (GFT).} 
		\label{fig:adjoint}
\end{figure}

\subsubsection*{Adjoint representation}

As discussed in Section~\ref{sec:setup}, the OPE $Q\times Q$ contains again operators in the adjoint representation.  This offers the possibility to test the effect of a $\mathbb Z_2$ symmetry in the CFT. If $Q$ is odd under such a symmetry, then the three-point function $\langle QQQ\rangle$ must vanish. When inputting gaps on the adjoint scalar sector above the external dimension $\Delta_Q$, we then have the choice to allow the presence of an isolated contribution with $\Delta=\Delta_Q$  or to forbid it. This corresponds to imposing the presence of a $\mathbb Z_2$ symmetry under which $Q$ is  odd or not. Interestingly, for $N\geq4$, the bounds  in the two cases coincide. 

As shown in Figure~\ref{fig:adjoint}, a family of defined kinks arises for large values of $N$. The location of the kink in $\Delta_Q$ moves to the left as $N$ increases but it does not seem to converge to the free value $1/2$. The inspection of the extremal functional at the kink for the case $N=100$ displays a behavior similar to the kink explored in the Ising model in \cite{El-Showk:2014dwa}: the second scalar adjoint operator decouples from the left, while the spin-2 operator after the stress tensor decouples from the right. 

In principle one could create an island inputting gaps in these two sectors. We leave this exercise for the future, as at present we do not have a candidate CFT for this family of kinks.
From Figure~\ref{fig:adjoint} we can speculate that this family of putative CFTs must have a $\mathbb Z_2$ symmetry under which $Q$ is odd. The small values of $\Delta_Q$ disfavor gauge theories, especially because at large-$N$ anomalous dimensions are suppressed, see section~\ref{sec:gaugetheories}. We also saw in section~\ref{sec:LGW} that a simple theory of scalars does not have fixed points for $N\gtrsim4$: it would be interesting to systematically search for fixed points of  Gross-Neveu-Yukawa like theories theories involving scalars in the adjoint  (even in presence of supersymmetry) and compare them with our results.

\begin{figure}[thbp]
\includegraphics[width=0.48\textwidth]{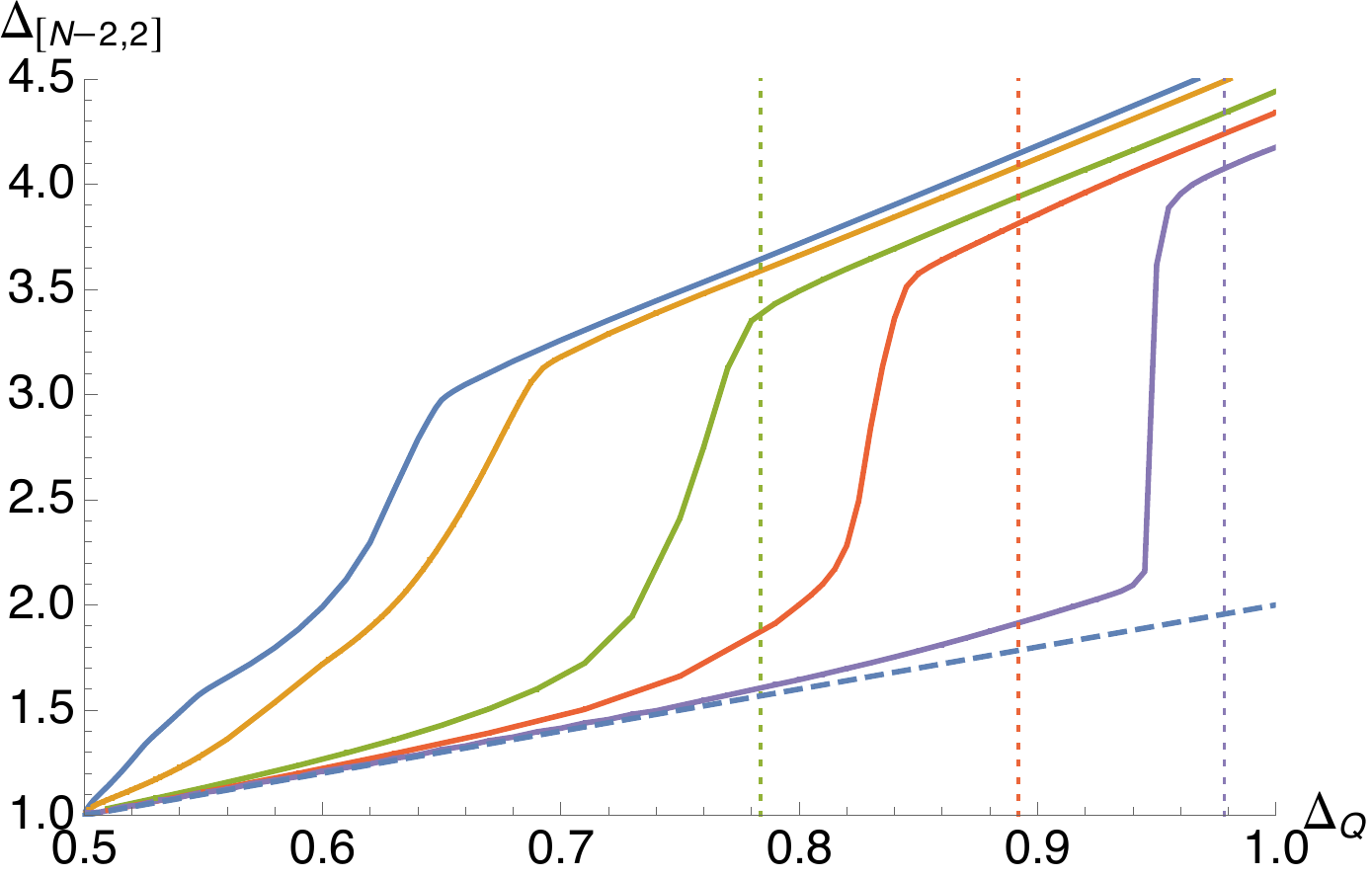}
\includegraphics[width=0.48\textwidth]{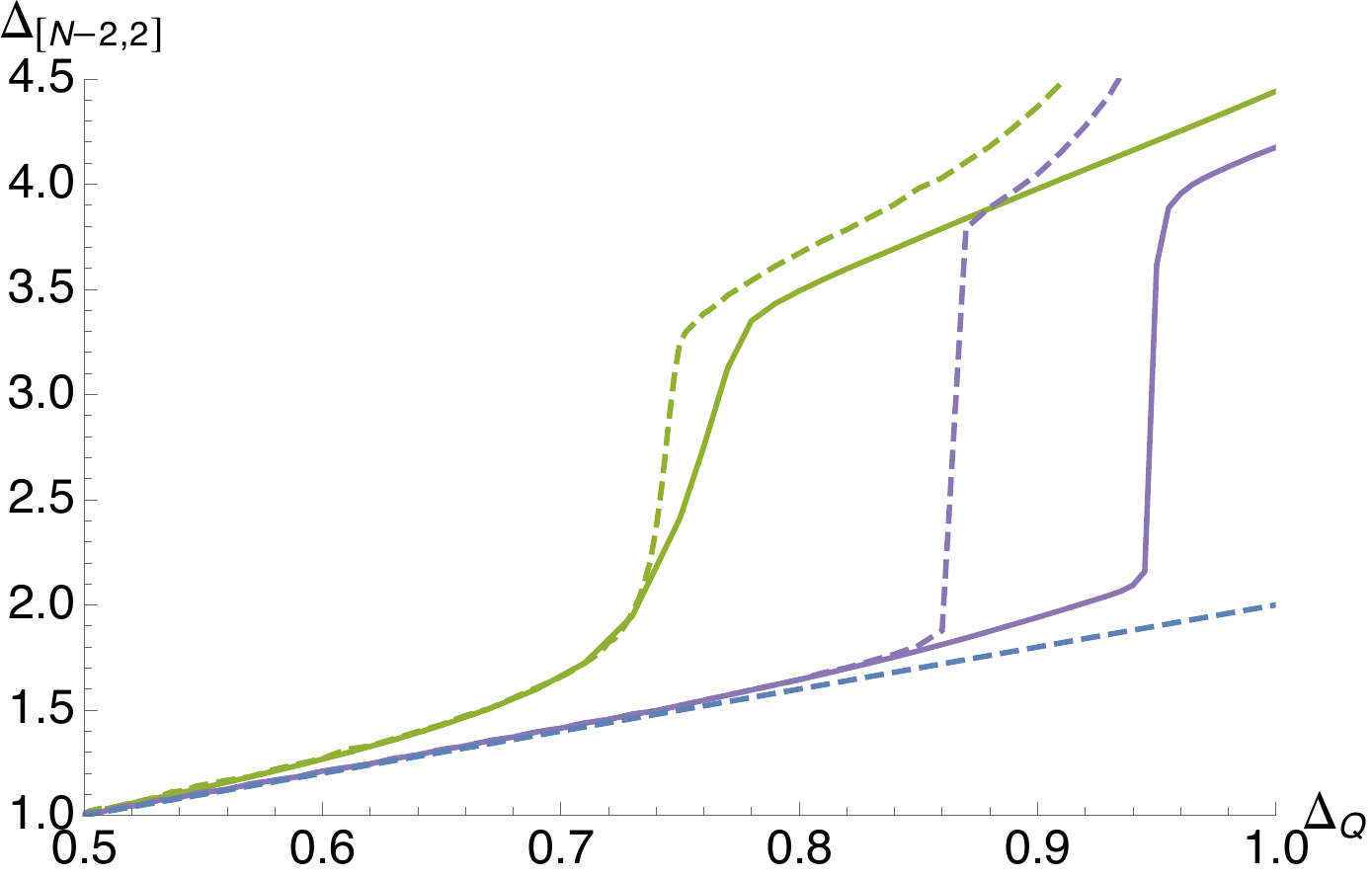}
		\caption{Bounds on the dimension of the first scalar in the $[N-2,2]$ representation \descriptionNColors  For $N=4$ there are kinks at $\Delta_{t}=0.54$ and $\Delta_{t}=0.64$ respectively. For larger $N$ the first kink disappears. A family of sharp kinks is visible for all $N$. Green, red and purple vertical dotted lines show the large-$N$ values of $\Delta_Q$ for bosonic QED${}_3$ with $N=10,20,100$.} 
		\label{fig:bisectionallnoverviewbox}
\end{figure}

\subsubsection*{Representation $[N-2,2]$}

Next we examine the bound on the dimension of the first scalar  operator in the representation $[N-2,2]$, see Figure~\ref{fig:bisectionallnoverviewbox}. These bounds display a second family of very sharp kinks for all $N$, moving to the right towards $\Delta_Q=1$ as $N$ increases.  In order to obtain the correct scaling with $N$ it was crucial to impose that $Q$ is actually the lightest adjoint scalar in the theory.\footnote{We are grateful to authors of \cite{Su:2021ToAppear} for discussing this point.} One can always restrict to this case with no loss of generality. As shown in the right panel of Figure~\ref{fig:bisectionallnoverviewbox}, this assumption proved very effective, especially for large $N$. As discussed in Section~\ref{sec:gaugetheories}, the lightest scalar operator in this representation in abelian gauge theories is expected to have dimension $4+O(1/N)$, while in LGW models or non-abelian gauge theories we expect operators with lower dimensions.  
Hence, in the left panel we compare with the prediction for tricritical $\mathrm{QED}_3$ of \cite{Benvenuti:2019ujm}.\footnote{The prediction for $CP^{N-1}$ has larger $\Delta_Q$.} For $N=100$, where we expect to trust the large-$N$ expansion, the kink is still a bit far away. The discrepancy gets larger for $N=20$. Incidentally, the kink at $N=10$ coincides exactly with the large-$N$ prediction. We interpret this as a coincidence, since we do not expect the prediction to be reliable for this value of $N$. Indeed, as we will see in the next subsection, bootstrap bounds are in tension with the large-$N$ prediction for $N=10$.\footnote{In other cases bootstrap bounds have been found to disagree with large-$N$ prediction even for large values of $N$ \cite{Bissi:2020jve}.}

\begin{figure}[thbp]
\includegraphics[width=0.48\textwidth]{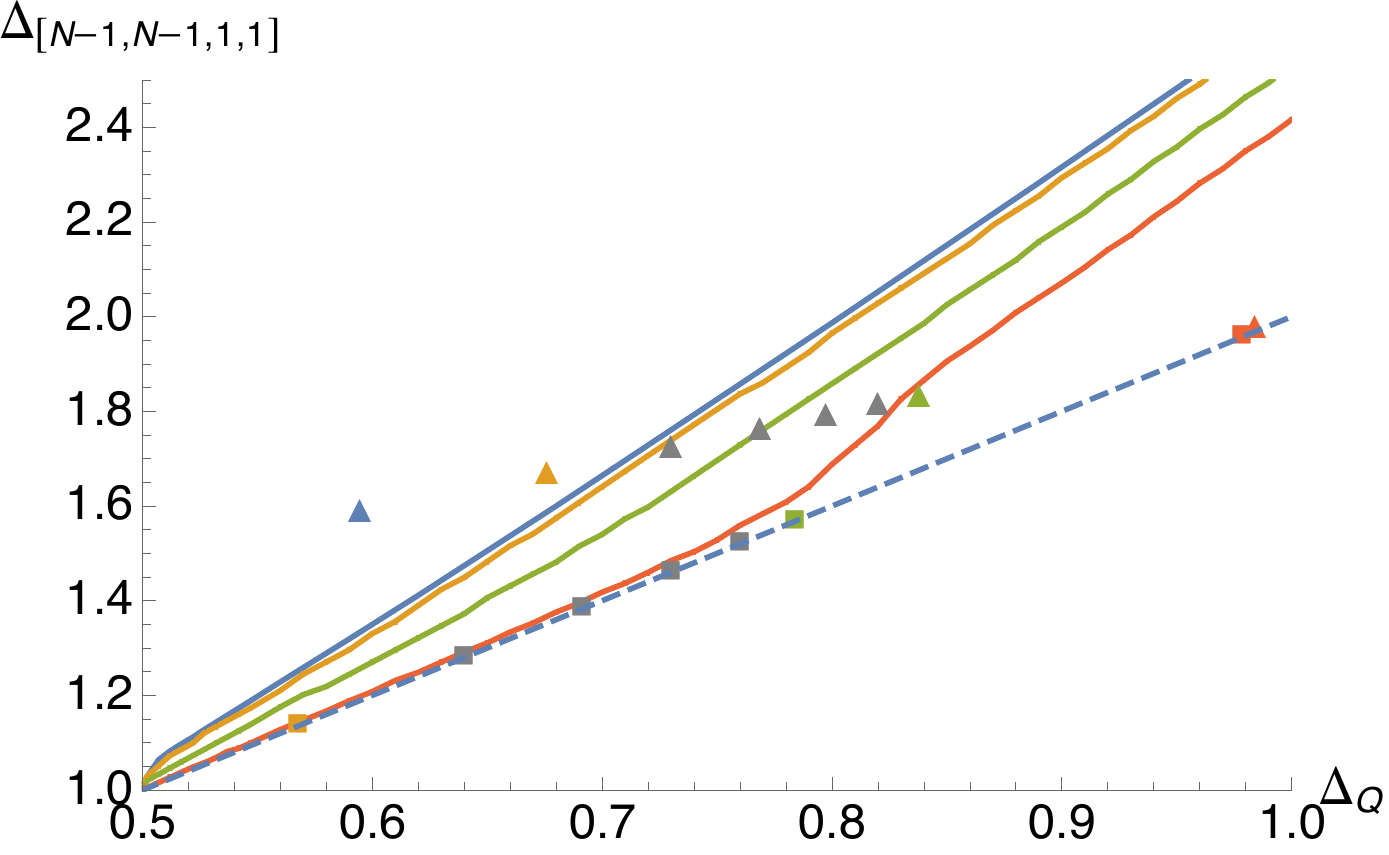}
\includegraphics[width=0.48\textwidth]{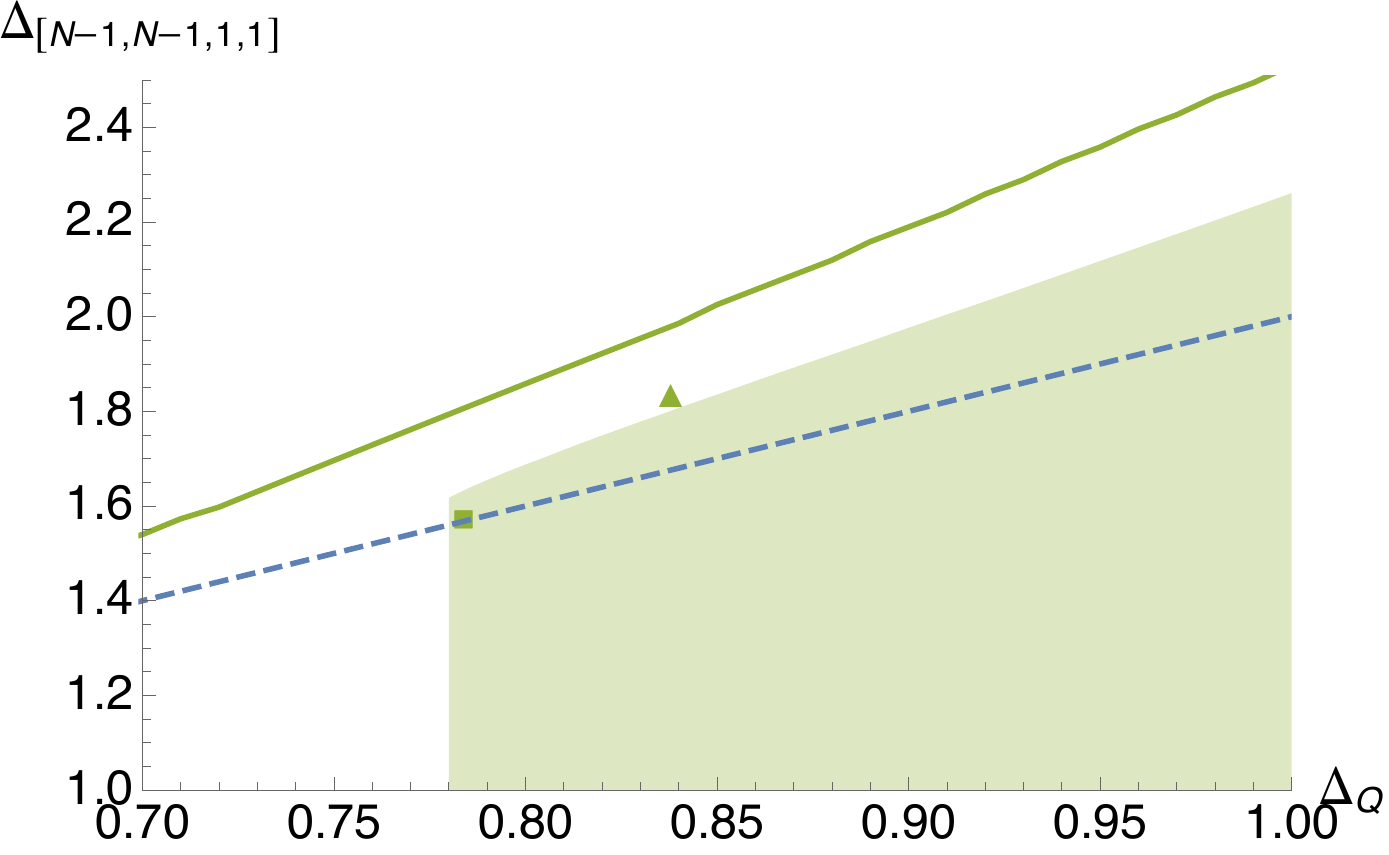}
		\caption{Bounds on the dimension of the first scalar in the $[N-1,N-1,1,1]$ representation. \descriptionNColorsMinus The bounds have been obtained at $\Lambda=19$. The triangles and squares are, respectively, Abelian Higgs and bosonic QED conformal dimensions in \eqref{eq:biadjoint} for $N=4,5,6,7,8,9,10,100$.} 
\label{fig:bisectionallnoverviewv31}
\end{figure}

\subsubsection*{Representation $[N-1,N-1,1,1]$}

We conclude with the fourth scalar representation. As shown in Figure~\ref{fig:bisectionallnoverviewv31} the bounds do not reserve particular surprises. Zooming on the region close to the unitarity bound one would observe a series of kinks corresponding to the dimension of traceless symmetric tensor in the $O(N')$ vector model, which is again a special solution of our crossing equations. The same kinks would also be visible in Figure~\ref{fig:bisectionallnoverviewbox}. 

Despite not showing any other special feature, these bounds offer the possibility to test the validity of the large-$N$ expansion in the Abelian Higgs model. The leading operator in the representation $[N-1,N-1,1,1]$ is made of four scalars and its dimension has been computed in the case of bosonic QED${}_3$ and in the case of Abelian Higgs model \cite{Benvenuti:2019ujm}. While in the former $\Delta_{[N-1,N-1,1,1]}=2\Delta_Q$, in the latter che predictions are not always in the allowed region. As expected, Figure~\ref{fig:bisectionallnoverviewv31} shows that for $N=4,5$ the large-$N$ prediction is noticeably wrong. On the contrary $N\geq10$ is slightly inside the allowed region. One can wonder if additional assumptions allow to get closer the predicted point. According to \cite{Benvenuti:2019ujm}, for $N=10$ one has $\Delta_S\simeq1.838$. Thus, in the right panel of Figure~\ref{fig:bisectionallnoverviewv31}, we assumed a conservative gap $\Delta_S\geq1.5$, $\Delta_{[N-2,2]}\geq3$, $\Delta_{\mathrm{adj}}\geq\Delta_Q$ and computed again the bound on $\Delta_{[N-1,N-1,1,1]}$. The result does not show any feature, but excludes the large-$N$ leading order prediction. The fact that bosonic QED${}_3$  seems to be at the corner is a pure coincidence, since this theory is excluded by our assumptions.
We tried several gaps, but the allowed region depends smoothly on them, with no striking evidences of CFTs.


\section{\texorpdfstring{Focusing on $\boldsymbol{SU(4)}$}{Focusing on SU(4)}}
\label{sec:resultsSU4}

\begin{figure}[ht]
	\centering
	\includegraphics[width=0.45\textwidth]{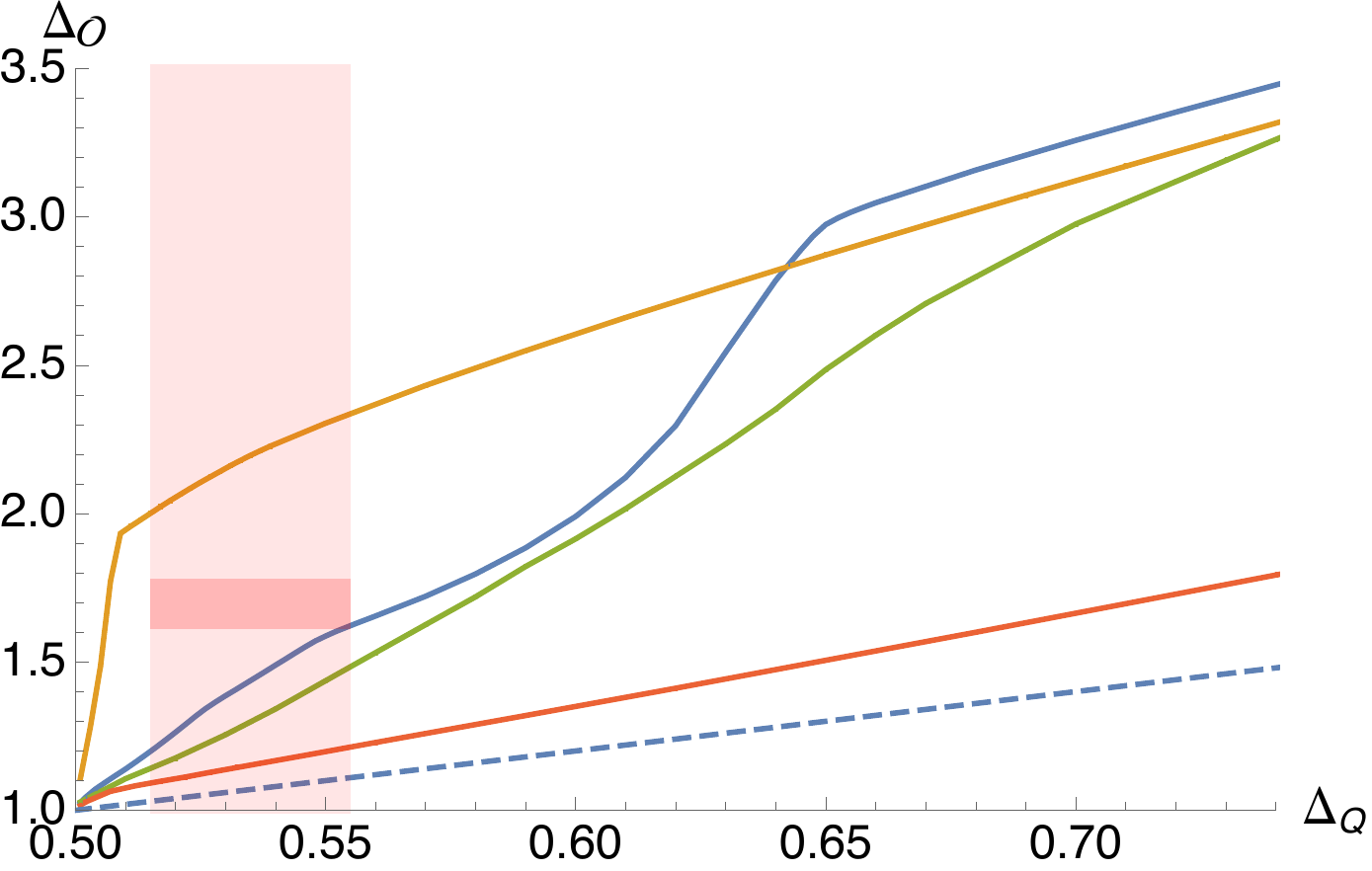}
	\includegraphics[width=0.45\textwidth]{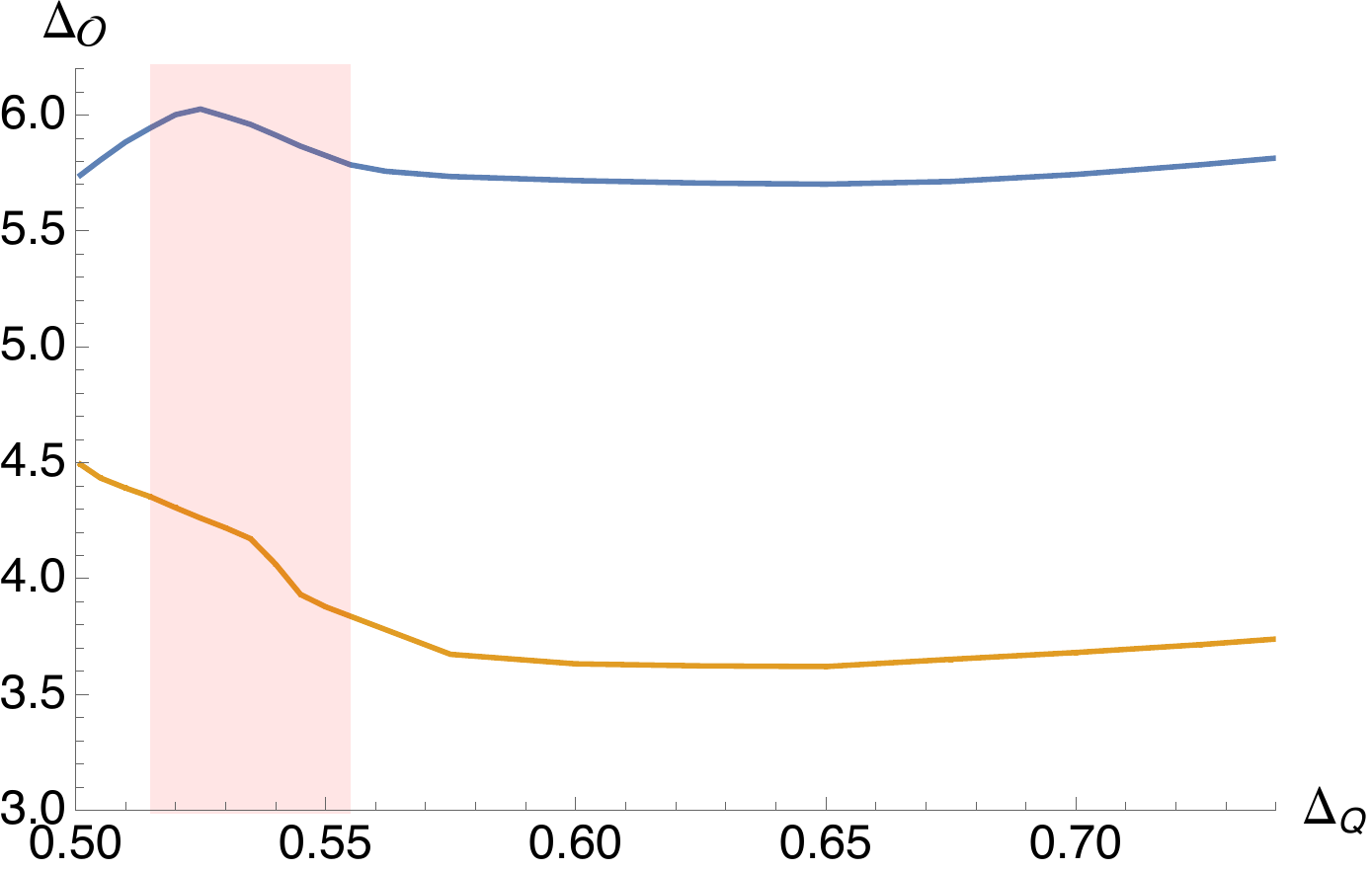}
	\caption{Bounds on the dimension of various operators in $SU(4)$. On the left: The blue, orange and green and red lines correspond respectively to the bond on the dimension of the leading scalar $[N-2,2]$, singlet, adjoint  and $[N-1,N-1,1,1]$ irreps. On the right: bound on the dimension of  the first spin-2 operator after the stress tensor (blue) and first spin-1 operator after the adjoint $SU(4)$ current. The bounds have been computed at $\Lambda=19$. The light band corresponds to the prediction of lattice simulation for $\Delta_Q$. The darker square shows the lattice prediction for the singlet dimension as well.}
	\label{fig:summaryBoundSU4}
\end{figure}

Let us now focus on the case $N=4$. For this specific case there is a well defined candidate CFT to compare with. This is the $ACP^3$ lattice model studied in \cite{Pelissetto:2017sfd}. Using 
 lattice simulation they  find a fixed point with an adjoint scalar of dimension $\Delta_Q=0.535\pm0.02$ and exactly one relevant singlet with dimension $1.69\pm0.8$. Our goal is to verify if the boostrap bounds show any evidence of such CFT.

\begin{figure}[h!]
	\centering
	\includegraphics[width=0.55\textwidth]{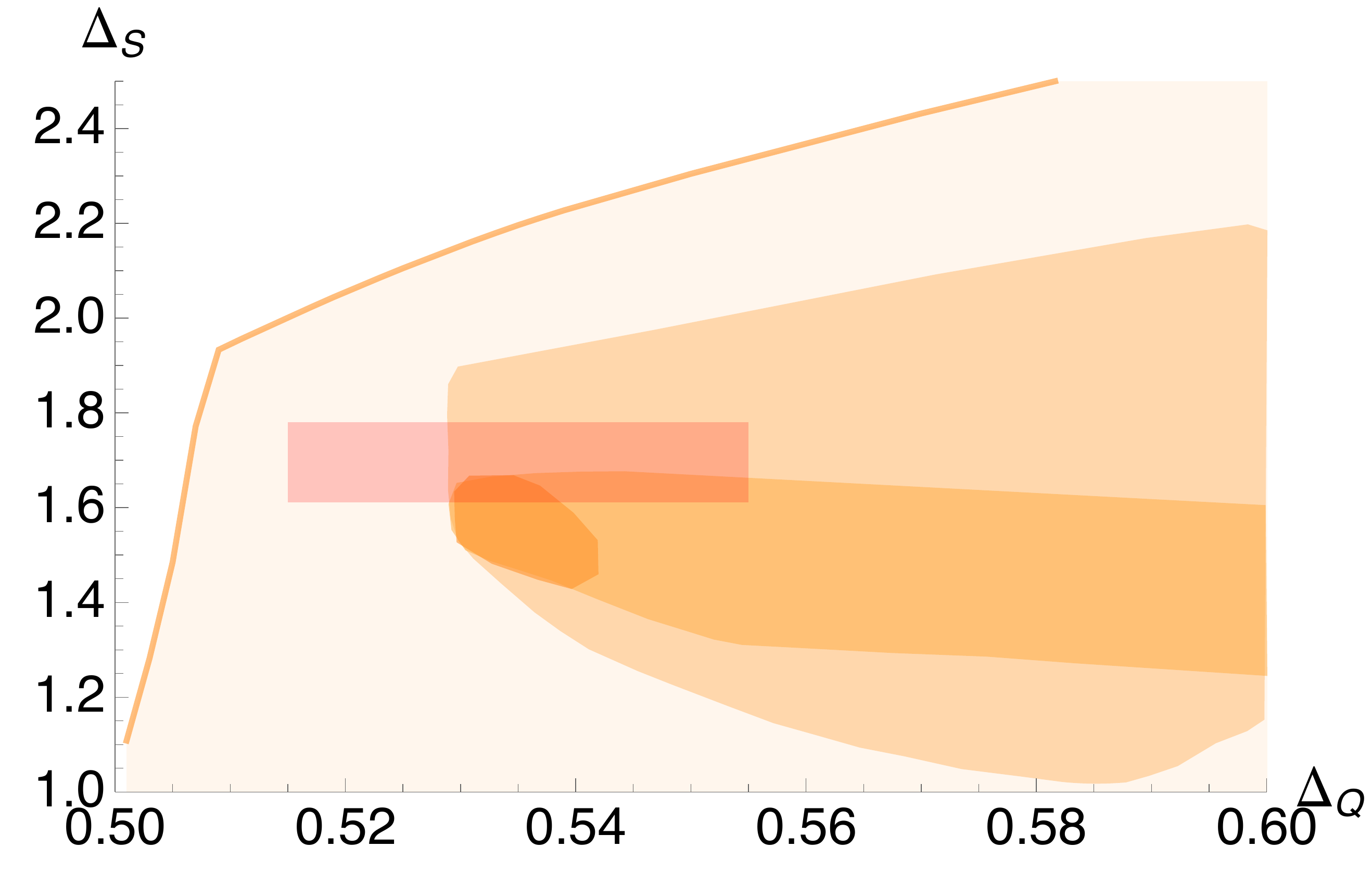}
	\caption{Allowed regions in the plane $(\Delta_Q,\Delta_S)$, with increasingly strong assumption. The assumptions are (from light to dark): no assumptions; $S$ is the only relevant scalar and $\Delta_{J'}\geq3$ and $\Delta_{T'}\geq 4,5.5,5.7$. As usual $J'$ denotes the next spin-1 operator after the adjoint current and $T'$ denotes the next spin-2 operator after the stress tensor. The red box is the lattice prediction for $ACP^3$. The bounds have been computed at $\Lambda=19$.}
	\label{fig:TprimeBoundSU4}
\end{figure}

In Figure~\ref{fig:summaryBoundSU4} we report the bounds for scalar and non scalar operators for the specific case of $N=4$ and compare them with the lattice determinations. The bound on $T'$, the first spin-2 operator after the stress tensor, and $J'$ the first spin-1 operator after the adjoint current, look promising; in addition the representation $[N-2,2]$ has a milder kink in the region of interest. The situation looks remarkably similar to the case of $ARP^3$ studied in  \cite{Reehorst:2020phk}. Also in this case one can check that imposing increasing  gaps on $\Delta_{T'}$ allows to carve out  islands in the $(\Delta_Q,\Delta_S)$ plane that shrink around the lattice prediction, see Figure~\ref{fig:TprimeBoundSU4}. Unfortunately in \cite{Reehorst:2020phk} a bootstrap study of mixed correlators involving a scalar singlet operator revealed that part of the peak in the $\Delta_{T'}$ bound was excluded. Nevertheless island overlapping with lattice prediction could be created by imposing suitable gaps in other channels. We leave a detailed investigation of this direction for a future work.

We conclude noticing that the scalar bounds in Figure~\ref{fig:summaryBoundSU4} impose the presence of relatively small dimension operator beside the singlet, in particular \mbox{$\Delta_{[N-1,N-1,1,1]}\lesssim 1.21$} in the region of interest. It would be extremely interesting to estimate $\Delta_{\mathrm{adj}}$, $\Delta_{[N-1,N-1,1,1]}$  and $\Delta_{[N-2,2]}$ in $ACP^3$ on the lattice and compare with our bounds.

\begin{figure}[h!]
	\centering
	\includegraphics[width=0.5\textwidth]{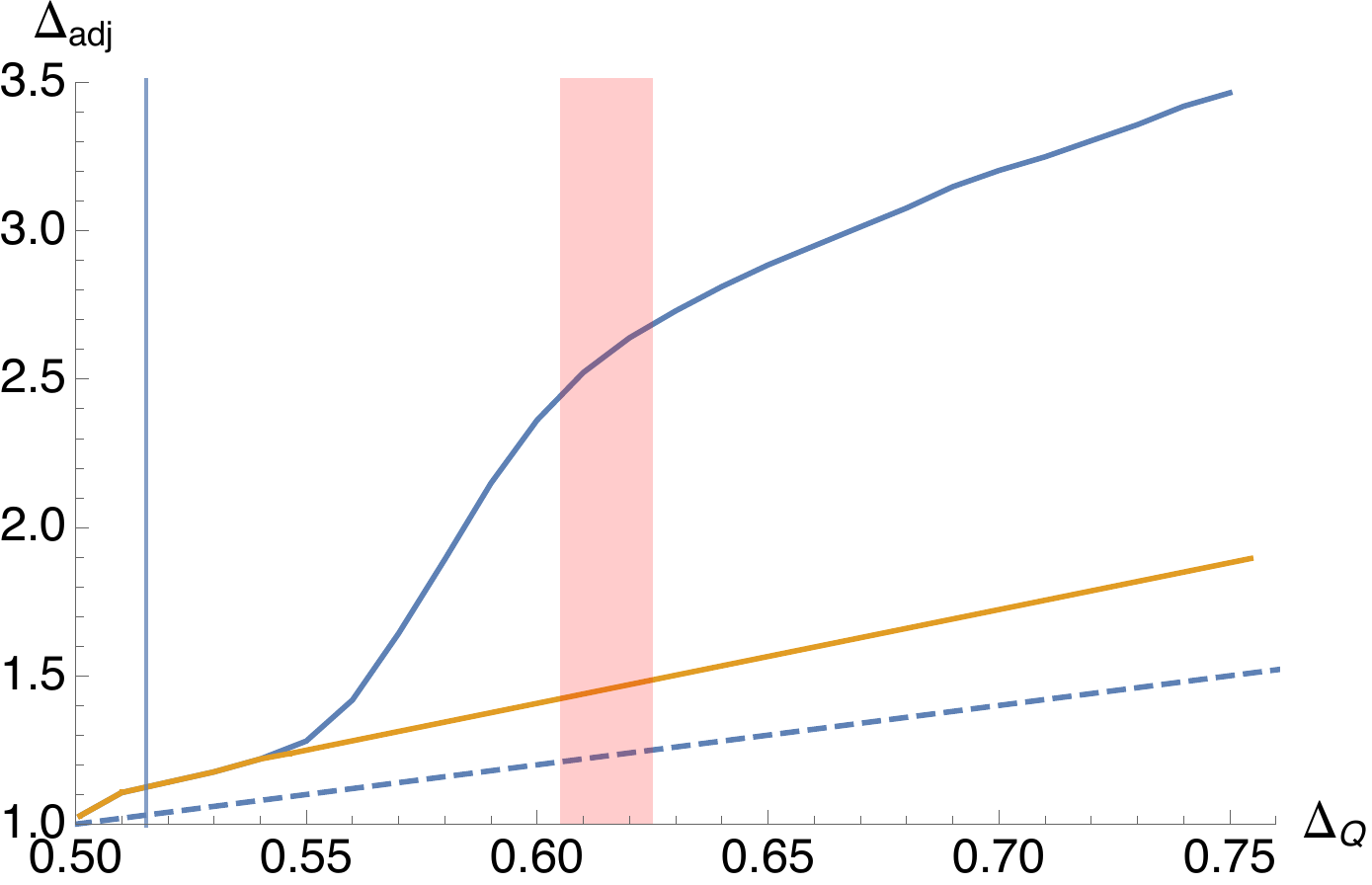}
	\caption{Bounds on the first adjoint operator for the case of $SU(3)$, with the assumption that $Q$ appears (blue) or does not appear (yellow) in its own OPE  $Q\times Q$. The dashed line corresponds to GFT solution. The bounds have been computed at $\Lambda=27$. The light red band corresponds to the prediction of the loop model. The vertical line corresponds to the $O(8)$ large-$N$ prediction.}
	\label{fig:SingletBoundSU3}
\end{figure}

\section{\texorpdfstring{Focusing on $\boldsymbol{SU(3)}$}{Focusing on SU(3)}}
\label{sec:resultsSU3}

Finally let us consider the case of $N=3$. As discussed in section~\ref{sec:LatticeAH}, there are lattice evidences of a second order phase transition, corresponding to the ferromagnetic $CP^2$ model. 

From the group theory point of view, this case is special, since the $Q\times Q$ OPE exchanges one representation less. Let us recall the group theory decomposition in this case:  
\be
\bf{8} \otimes \bf{ 8} = \bf{1}_+ \oplus \bf{8}_+ \oplus \bf{8}_-  \oplus \bf{27}_+ \oplus (\bf{10} + \bf{\overline{10}})_-\,,
\ee
where in the notation of the previous section one has $[N-1,1]\rightarrow \bf 8$, $[N-2,1,1]\rightarrow \bf 10$ and \mbox{$[N-1,N-1,1,1]\rightarrow \bf 27$}. The representation $[N-2,2]$ is missing.

Among the bound on scalars, we only show the bound on the first adjoint scalar, with and without assuming the existence of a $\mathbb Z_2$ symmetry under which $Q$ is odd. As shown in Fugure~\ref{fig:SingletBoundSU3} the two bounds coincide for $\Delta_Q\lesssim 0.54$ and the differ substantially: if $Q$ appears in the OPE $Q\times Q$, the bound on the next scalar is much higher. In particular the bound displays a rounded kink in the region predicted by the loop model discussed in section~\ref{sec:LatticeAH}. This is interesting since in the ferromagnetic case the order parameter is even under exchange of two sublattices, contrarily to the antiferromagnetic case, see discussion in \cite{Pelissetto:2017sfd}.

The bound on the singlet looks exactly as Figure~\ref{fig:bisectionallnoverviewsinglet}, displaying only a sharp kink corresponding to the $O(8)$ model. The loop prediction for $\Delta_Q$ and $\Delta_S$ lies well inside the allowed region. The bound on $\Delta_{27}$ is instead smooth and slightly above the GFT line.

\begin{figure}[h!]
	\centering
	\includegraphics[width=0.45\textwidth]{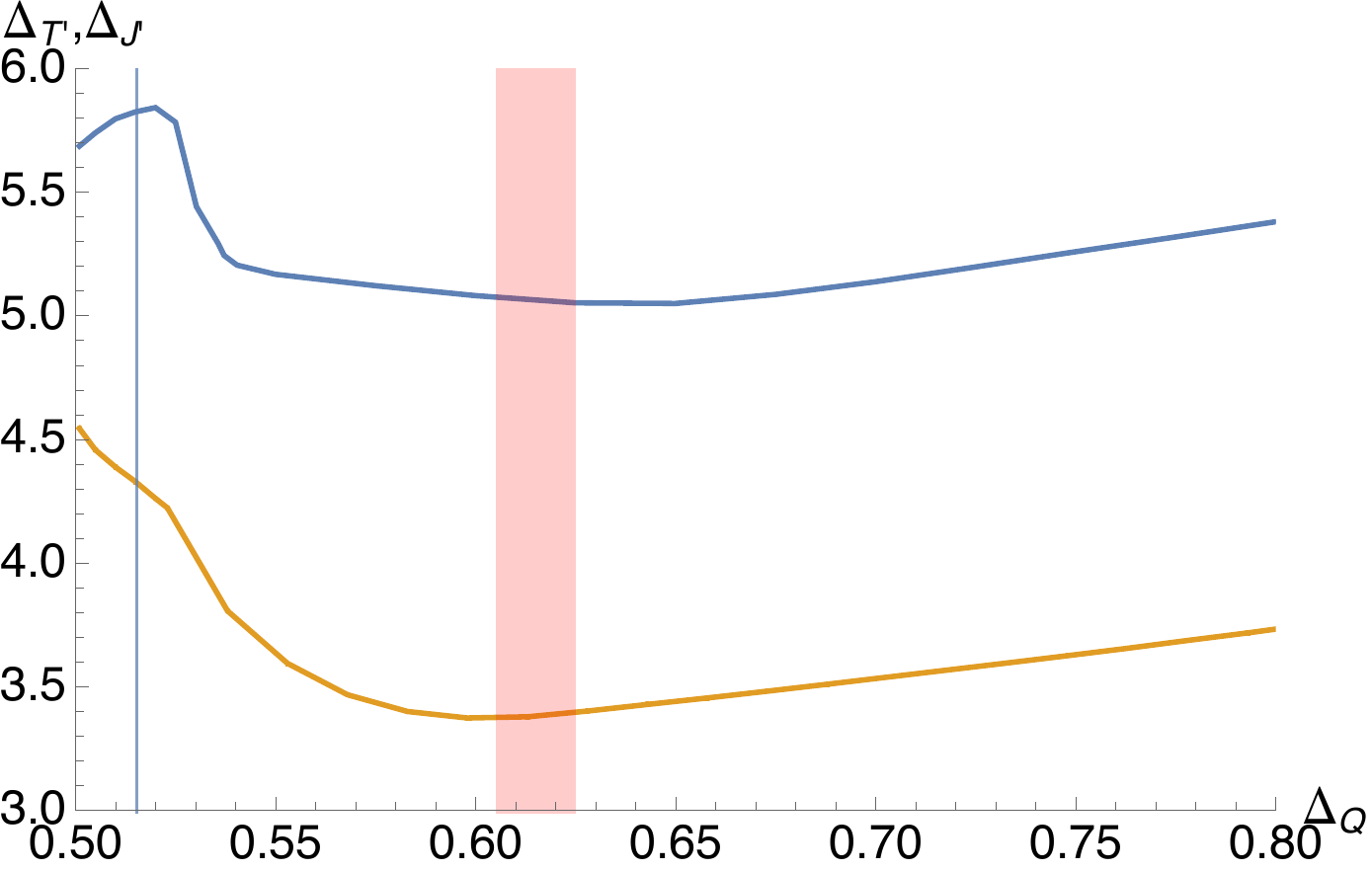}
	\includegraphics[width=0.45\textwidth]{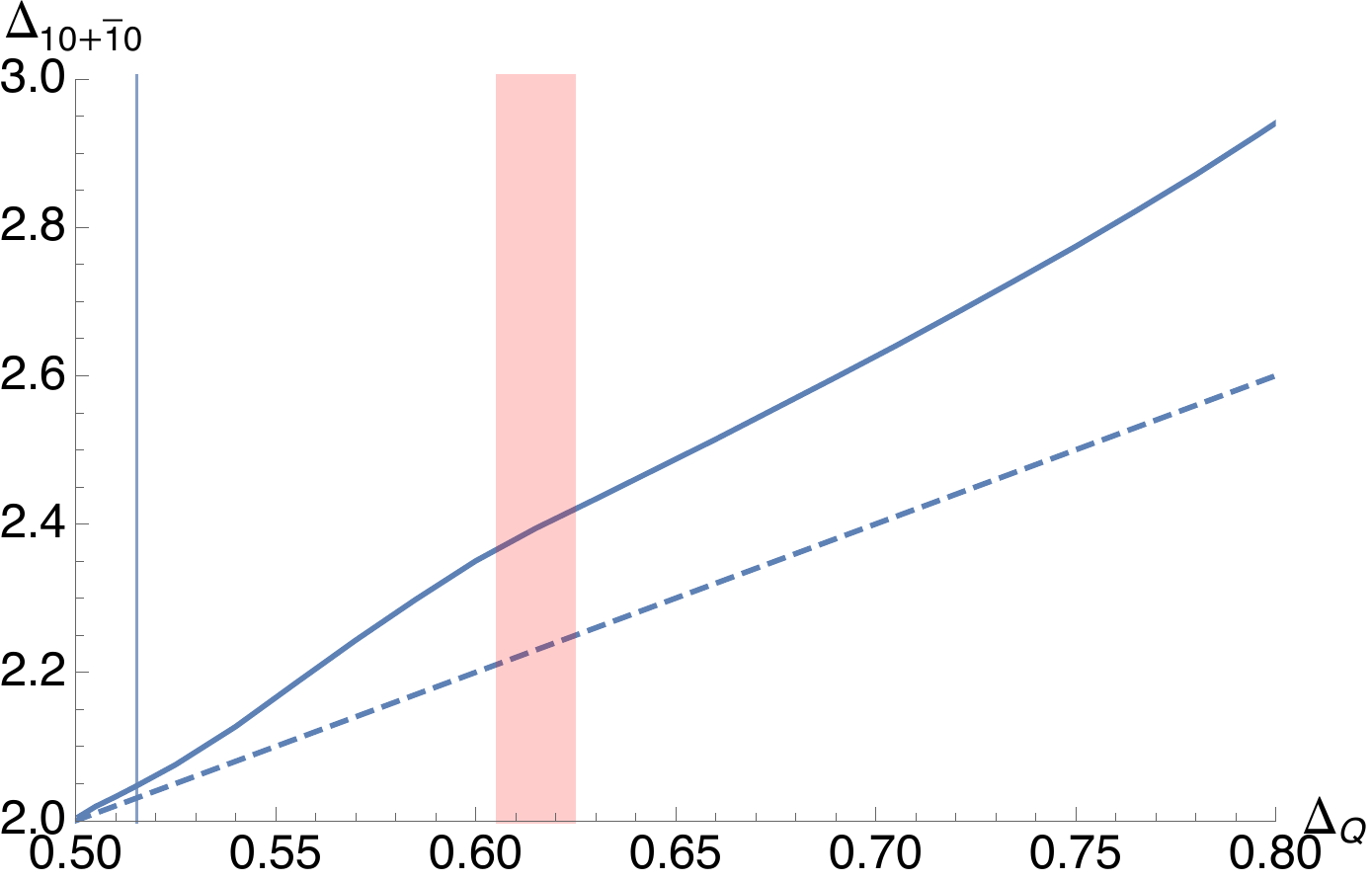}
	\caption{Bounds on the dimension of various spinning operators in $SU(3)$. On the left: The blue, orange and green lines correspond respectively to the bond on the dimension of: the first spin-2 operator in the singlet sector after the stress tensor; the first spin-1 operators in the adjoin sector after $SU(3)$ conserved current. On the right: bound on the dimension of  the first spin-1 operator in the representation $\bf{10+\overline{10}}$. The bounds have been computed at $\Lambda=19$. The light red band corresponds to the prediction of the loop model. The vertical line corresponds to the $O(8)$ large-$N$ prediction.}
	\label{fig:summarySU3}
\end{figure}

In Figure~\ref{fig:summarySU3} we show instead bounds on spinning operators. The operators after the stress tensor and after the adjoint current have features corresponding to the $O(8)$ model only. The other representation instead has a mild kink in correspondence of the loop model, see right panel of Figure~\ref{fig:summarySU3}. 

\begin{figure}[h!]
	\centering
	\includegraphics[width=0.45\textwidth]{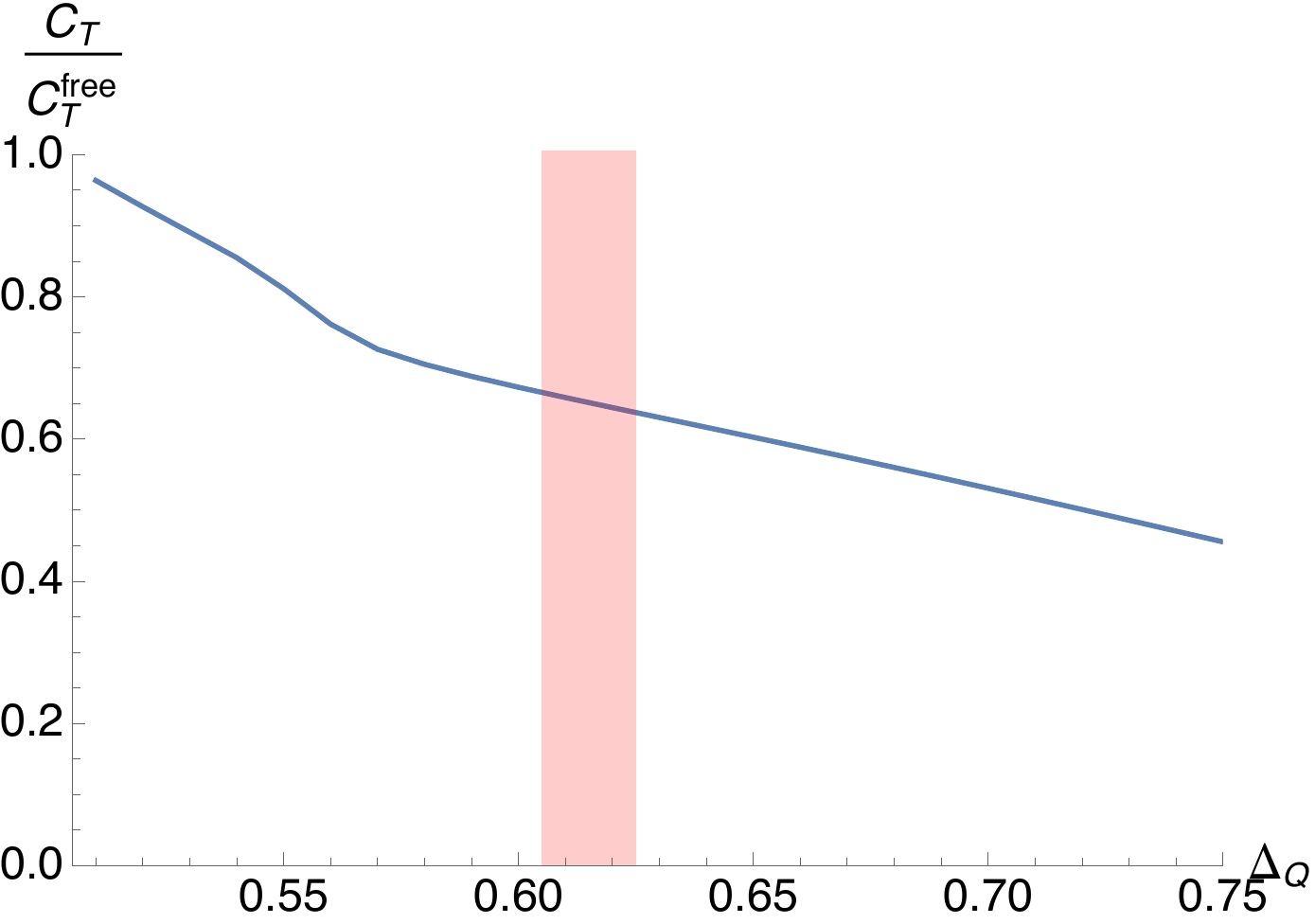}
		\includegraphics[width=0.45\textwidth]{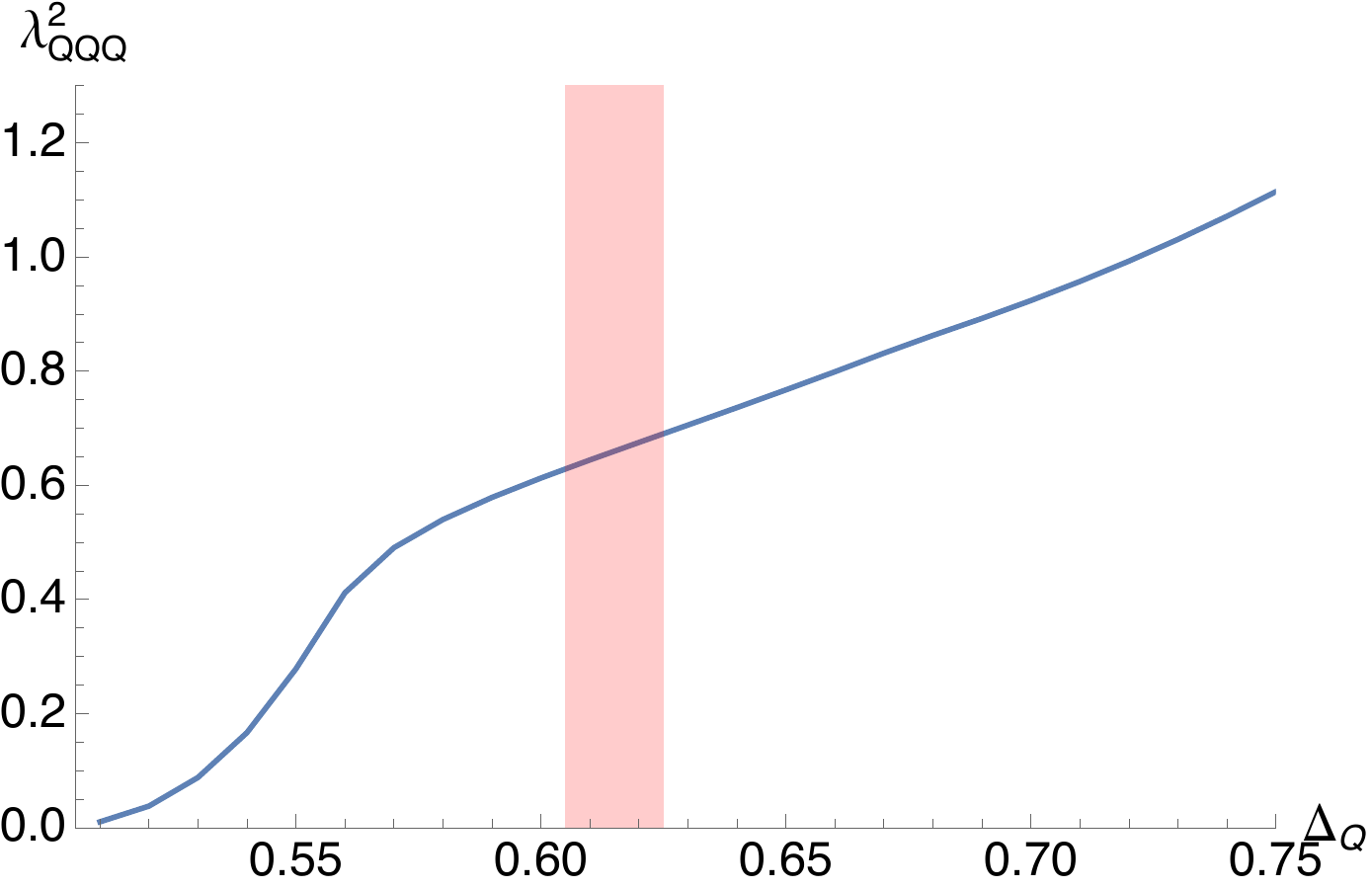}
	\caption{On the left: lower bounds on the central charge $C_T$ in  $SU(3)$, normalized to the free value for 8 real scalars. On the right: upper bounds on the OPE coefficient $\lambda^2_{QQQ}$ in $SU(3)$. The bounds have been computed at $\Lambda=27$. The light red band corresponds to the prediction of the loop model. Both bounds display a change of slope at $\Delta_Q\simeq 0.57$.}
	\label{fig:opeSU3}
\end{figure}

Finally we also explored bounds on the central charge $C_T$ and the OPE coefficient $\lambda_{QQQ}^2$, as shown in Figure~\ref{fig:opeSU3}, which instead display a change of slope for $\Delta_Q\sim 0.57$, outside of the loop model prediction.

It would tempting to interpret the  features observed in Figures~\ref{fig:SingletBoundSU3} and \ref{fig:summarySU3} as signals of a CFT and identify it with the second order phase transition observed in \cite{Nahum:2013qha}. As a further evidence, the extremal functional spectrum along the bound in Figures~\ref{fig:SingletBoundSU3}, for $\Delta_Q\sim 0.61$, has a unique singlet operator inside the interval predicted by the loop model, $\Delta_S\in [1.08,1.18]$.  

\begin{figure}[h!]
	\centering
	\includegraphics[width=.8\textwidth]{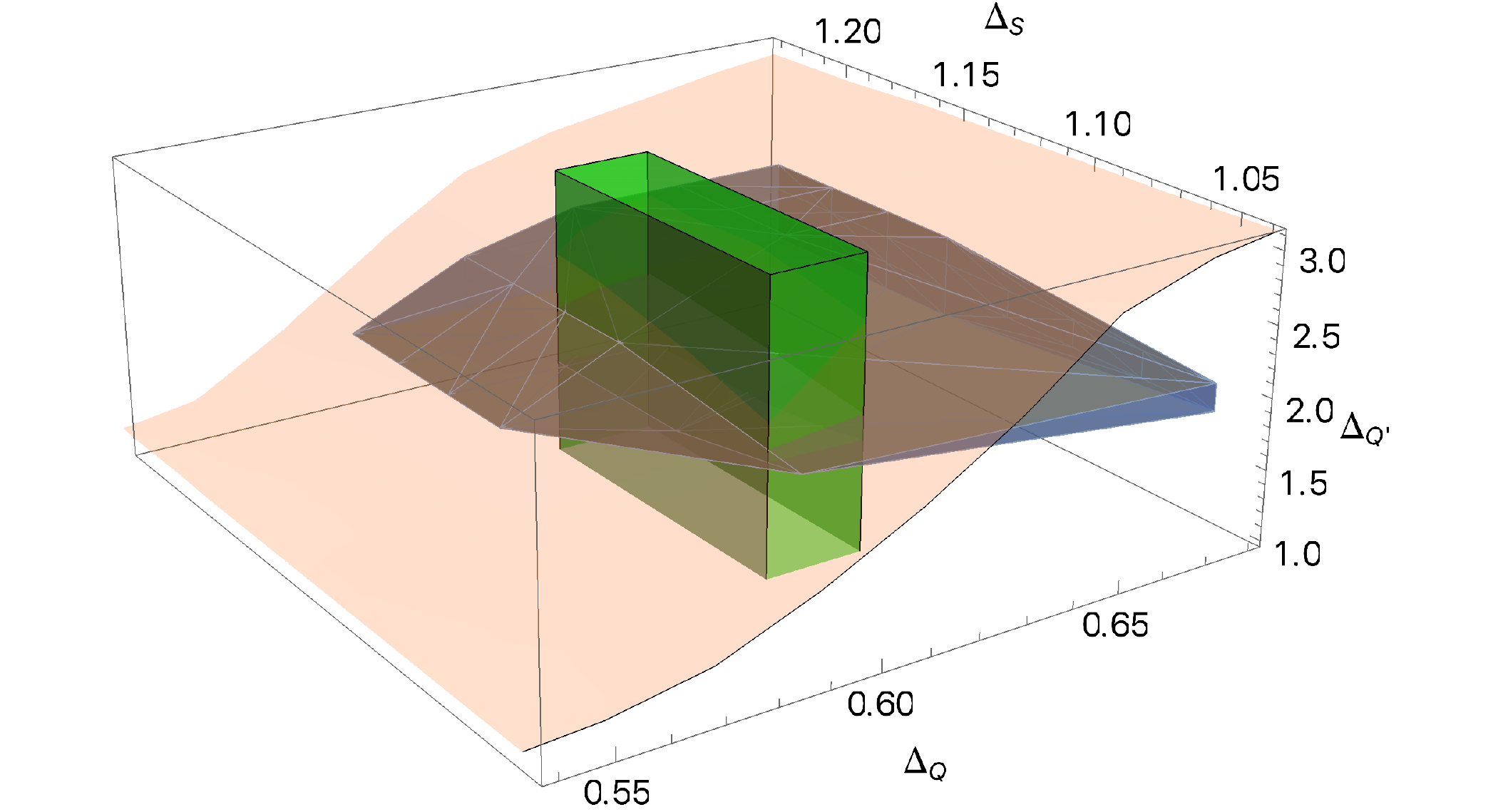}
	\caption{Allowed region in the space $(\Delta_Q, \Delta_{Q'}, \Delta_S)$, the dimensions of the first two adjoint scalars $Q,Q'$ and the first singlet scalar $S$, assuming that all other adjoint and singlet scalars are irrelevant. The region has been obtained imposing crossing symmetry in the mixed system of correlators involving $Q$ and $S$. The orange shaded surface represents the upper bound on $\Delta_{Q'}$ from the single correlator of $Q$. The green region is the prediction of the loop model. The blue allowed region has been computed at $\Lambda=19$.}
		\label{fig:mixedSU3}
\end{figure}

Although we tried many different analyses, we have been unable to create a closed region in the $(\Delta_Q,\Delta_S)$ plane. We also tried to study the mixed system of correlation functions involving the operator $Q$ and the first scalar singlet $S$. In Figure~\ref{fig:mixedSU3},  we show a scan in the three dimensional space $(\Delta_Q,\Delta_S,\Delta_{Q'})$, assuming they are the only relevant scalar operators in the singlet and adjoint sector (and no extra $\mathbb Z_2$).\footnote{More precisely we assumed a gap in the adjoint sector $\Delta_{Q''}\geq 3.5$, both in the $Q\times Q$ and $Q\times S$ OPEs.} We find an allowed region consistent with the lattice prediction. In the region of interest one must then have a second relevant adjoint scalar with dimension $\Delta_{Q'}\in [1.6,2.2]$. 

\section{Conclusions}
\label{sec:conclusions}

In this work we continue the exploration of the space of CFTs, focusing on theories with a scalar operator $Q^b_a$ transforming in the adjoint representation of a global $SU(N)$ symmetry. We concentrate our attention to relatively small scalar dimensions $\Delta_Q\lesssim 1$, as our main target are gauge theories with complex scalar matter, as well as $CP^{N-1}$ and related models. Gauge theories with fermionic matter are indeed expected to have larger dimension operators and their bootstrap analysis is generically harder, see \cite{Li:2018lyb,Li:2020bnb} for recent attempts.

 A systematic study of the correlation function $\langle QQQQ\rangle$ for general $N$ revealed new features and confirmed the importance of bounding non-singlet representations. Most notably, a family of sharp kinks appears for all $N\geq4$ in the bound on the first scalar $\mathcal O^{[a \, b]}_{[c\, d]}$ with two pairs of antisymmetric indices (we called this representation $[N-2,2]$ in the main text). Due to the antisymmetric properties of the indices, this operator is expected to be heavy, $\Delta\sim 4$, in abelian gauge theories and in $CP^{N-1}$ models, while non-abelian gauge theories, GFTs and LGW theories can contain smaller dimension scalar with $\Delta\lesssim 2 $. It would be tempting to identify these kinks, shown in Figure~\ref{fig:bisectionallnoverviewbox}, with the fixed points of bosonic QED${}_3$, however the large-$N$ predictions do not exactly match the bounds. Since these fixed points have been recently observed on the lattice \cite{Bonati:2020jlm}, it would be interesting to estimate the dimension of more scalar operators, with particular attention to the $[N-2,2]$ representation.
 
We also find a second family of kinks in the bounds on the first adjoint scalar appearing in the OPE $Q\times Q$. The kinks appear for large values of $N$ and they move closer to $\Delta_Q=1/2$ and $\Delta_S=1$ as $N$ increases; however, they do not seem to converge there, as shown in Figure~\ref{fig:adjoint}. The extremal functional spectrum reveals the presence of a stress tensor and a rearrangement of operators on the two sides of the kink similar to the Ising model case \cite{El-Showk:2014dwa}.  We do not have concrete candidates for this family of kinks: given the small values of $\Delta_Q$ it is unlikely that they correspond to a fixed point of gauge theories. Also, since theories of only adjoint scalars with quartic interactions do not posses fixed points at large-$N$, we speculate that these kinks could be associated to matrix models of scalars and fermions. 

Next, we focused on the case $N=4$, in the attempt to isolate a region corresponding to the phase transition observed in $ACP^3$ models by lattice simulations \cite{Pelissetto:2017sfd}. This case is very similar to the $ARP^3$ case discussed in \cite{Reehorst:2020phk}. We found several features supporting the existence of CFTs with critical exponents $\nu$ and $\eta$ compatible with the lattice predictions. In particular, by assuming a gap on the next spin-2 operator after the stress tensor, one can create a closed region overlapping with the lattice results. In order to get a more reliable and gap independent result, it will be important to bootstrap mixed correlators using the techniques introduced in  \cite{Chester:2019ifh,Chester:2020iyt}.

Finally, motivated by a second order phase transition observed in a loop construction of the compact ferromagnetic $CP^2$ model on the lattice \cite{Nahum:2013qha}, we studied the case of $N=3$.  In this case we found mild evidences of a putative CFT. Our analysis suggests that if such transition exists, the associated fixed point should not have a $\mathbb Z_2$ symmetry under which the order parameter $Q$ is odd: this is in agreement with the ferromagnetic construction. Moreover, the theory would contain a second relevant adjoint scalar with dimension $1.6 \lesssim \Delta_{Q'} \lesssim 2.2 $, as we show in Figure~\ref{fig:mixedSU3}. We believe a mixed correlator study of $Q,Q',S$ will help settling this question.

\section*{Acknowledgments}
We thank Slava Rychkov, Marten Reehorst, Andreas Stergiou, Ettore Vicari and Claudio Bonati for useful comments and discussions. For most of the duration of the project AM and AV were supported by the Swiss National Science Foundation under grant no.\ PP00P2-163670.  AM is also supported by Knut and Alice Wallenberg Foundation under grant KAW 2016.0129 and by VR grant 2018-04438. This project has received funding from the European Research Council (ERC) under the European Union's Horizon 2020 research and innovation programme (grant agreement no. 758903). All the numerical computations in this paper were run on the EPFL SCITAS cluster.

\appendix 
\allowdisplaybreaks

\section{Crossing equations for the adjoint and the singlet}

In this Appendix we show the crossing equations for the mixed correlator system the adjoint and the scalar representations of $SU(N)$. We will not assume a $\mathbb{Z}_2$ symmetry that sends $Q^a_b \to -Q^a_b$ in this setup, so the nonzero correlators are
\begin{equation}
\langle QQQQ\rangle\,,\qquad \langle QSSQ\rangle\,,\qquad \langle SSQQ\rangle\,,\qquad \langle QQQS\rangle\,,\qquad \langle SSSS\rangle\,,
\end{equation}
where $Q\in [N-1,1]$ and $S \in \bullet$. These equations were used to produce the plot in Figure~\ref{fig:mixedSU3}. 

\subsection{Four-point structures}

The four-point structures and crossing equations for $\langle QQQQ\rangle$ were discussed in Section~\ref{sec:setup}. Here we will list the additional structures arising in the correlators with $S$. Similarly as before, we define $\cT_4$ to be the only four-point structure of four scalars, but now with arbitrary external dimensions
\begin{equation}
\mathcal{T}_4 = \frac{1}{(x_{12}^2)^{\frac12(\Delta_A+\Delta_B)}(x_{34}^2)^{\frac12(\Delta_C+\Delta_D)}}\left(\frac{x_{24}^2}{x_{14}^2}\right)^{\frac12(\Delta_A-\Delta_B)}\left(\frac{x_{14}^2}{x_{13}^2}\right)^{\frac12(\Delta_C-\Delta_D)}\,,\label{eq:T4struct}
\end{equation}
where $\Delta_{A,B\ldots}$ is either $\Delta_Q$ or $\Delta_S$. Let us start with $\langle QSSQ\rangle$ and its permutation
\begin{subequations}
\begin{align}
\langle Q(\mathbf{p}_1)S(x_2)S(x_3)Q(\mathbf{p}_4)\rangle = \cT_4 \, (S_1\cdot\bar{S}_4)(S_4\cdot\bar{S}_1) \,h(u,v)
\,,\\
\langle S(x_1)S(x_2)Q(\mathbf{p}_3)Q(\mathbf{p}_4)\rangle = \cT_4 \, (S_3\cdot\bar{S}_4)(S_4\cdot\bar{S}_3) \,\tilde h(u,v)
\,.
\end{align}
\end{subequations}
In $h$ the representations exchanged are $[N-1,1]_\pm$ according to the parity of the spin and in $\tilde h$ only the singlet is exchanged. For the correlator with three $Q$'s we have two structures
\begin{equation}
\begin{aligned}
&\langle Q(\mathbf{p}_1)Q(\mathbf{p}_2)Q(\mathbf{p}_3)S(x_4)\rangle = \cT_4 \,\Bigl( \\&\qquad\qquad
\bigl((S_1\cdot\bar{S}_2)(S_2\cdot\bar{S}_3)(S_3\cdot\bar{S}_1) + (S_1\cdot\bar{S}_3)(S_2\cdot\bar{S}_1)(S_3\cdot\bar{S}_2)\bigr)\, h_+(u,v) \,+ \\&\qquad\qquad
\bigl((S_1\cdot\bar{S}_2)(S_2\cdot\bar{S}_3)(S_3\cdot\bar{S}_1) - (S_1\cdot\bar{S}_3)(S_2\cdot\bar{S}_1)(S_3\cdot\bar{S}_2)\bigr)\, h_-(u,v)
\Bigr)\,,
\end{aligned}
\end{equation}
where the representation exchanged in $h_\pm$ is the $[N-1,1]_\pm$. Finally the correlator $\langle SSSS\rangle$ has only one structure
\begin{equation}
\langle S(x_1) S(x_2)S(x_3)S(x_4)\rangle = \cT_4\, \tilde f(u,v)\,.
\end{equation}
and exchanges a singlet. This is all summarized in Table~\ref{tab:frOPEQS}.
\begin{table}
\centering
\begin{tabular}{lllll}
\toprule
function & $\cO$ exchanged & spin parity\\
\midrule
$\tilde f$ & $\bullet$  & even \\
$h_+$ & $[N-1,1]_+$ & even\\
$h_-$ & $[N-1,1]_-$ & odd\\
$h$ & $[N-1,1]_{(-1)^\ell}$  & any $\ell$\\
$\tilde h$ & $\bullet$ & even\\
\bottomrule
\end{tabular}
\caption{Operator contributions to the partial waves $\tilde f, h,\tilde h, h_\pm$.}\label{tab:frOPEQS}
\end{table}
Let us now write down the explicit conformal block expansions of $\tilde f$, $\tilde h$, $h$ and $h_\pm$
\begin{subequations}
\begin{align}
\tilde f(u,v) &= \sum_{\substack{\cO_{\Delta,\ell} \in \bullet\\\ell\;\text{even}}}\,(\lambda_{SS\cO}^{\bullet})^2 \,g^{0,0}_{\Delta,\ell}(u,v)\,, \\
\tilde h(u,v) &= \sum_{\substack{\cO_{\Delta,\ell} \in \bullet\\\ell\;\text{even}}}\,\lambda_{SS\cO}^{\bullet}\lambda_{QQ\cO}^{\bullet} \,g^{0,0}_{\Delta,\ell}(u,v)\,, \\
h(u,v) &= \sum_{\substack{\cO_{\Delta,\ell} \in [N-1,1]_+\\\ell\;\text{even}}}\,\bigl|\lambda_{QS\cO}^{[N-1,1]_+}\bigr|^2 \,g^{\delta,-\delta}_{\Delta,\ell}(u,v)+\sum_{\substack{\cO_{\Delta,\ell} \in [N-1,1]_-\\\ell\;\text{odd}}}\,\bigl|\lambda_{QS\cO}^{[N-1,1]_-}\bigr|^2 \,g^{\delta,-\delta}_{\Delta,\ell}(u,v)\,, \\
 h_+(u,v) &= \sum_{\substack{\cO_{\Delta,\ell} \in [N-1,1]_+\\\ell\;\text{even}}}\,\lambda_{QQ\cO}^{[N-1,1]_+}\lambda_{QS\cO}^{[N-1,1]_+} \,g^{0,\delta}_{\Delta,\ell}(u,v)\,, \\
 h_-(u,v) &= \sum_{\substack{\cO_{\Delta,\ell} \in [N-1,1]_-\\\ell\;\text{odd}}}\,\lambda_{QQ\cO}^{[N-1,1]_-}\lambda_{QS\cO}^{[N-1,1]_-} \,g^{0,\delta}_{\Delta,\ell}(u,v)\,.
\end{align}
\end{subequations}
For brevity we denoted $\delta \equiv \Delta_Q-\Delta_S$. The conformal blocks are normalized according to the convention expressed in~\eqref{eq:cbconvention}. That normalization implies $g_{\Delta,\ell}^{\delta,-\delta}\big(\frac14,\frac14\big)\geq0$ for all $\Delta,\ell,\delta$.

\subsection{Crossing equations}

We now need to put together the functions defined in the previous section with the functions $f_1,\ldots f_6$ defined in Subsection~\ref{ssec:fpf}. We also need to consider the special case where the operator exchanged in $h$ or $h_+$ is $Q$ itself or the operator exchanged in $\tilde f$ or $\tilde h$ is $S$ itself. In such a case we can use the permutation symmetry of the OPE coefficients $\lambda_{\cO_1\cO_2\cO_3} = \lambda_{\cO_3\cO_2\cO_1}$ to group the equations into a $3\times3$ matrix.

We define the following crossing symmetric combinations
\be
F_{\pm,\Delta,\ell}^{ABCD}(u,v) =  v^{\frac{\Delta_C+\Delta_B}{2}}\,g_{\Delta,\ell}^{\Delta_{AB},\Delta_{CD}}(u,v)\pm u^{\frac{\Delta_C+\Delta_B}{2}}\,g_{\Delta,\ell}^{\Delta_{AB},\Delta_{CD}}(v,u)\,.\label{eq:FHdefMixed}
\ee
The crossing equations then read
\begin{align}
(1\;\;1)\, V_{6;\lsp0,0}\left(\begin{matrix}1\\1\end{matrix}\right)=& \,
\big(\lambda_{QQQ}\;\;\lambda_{QSQ}\!=\!\lambda_{QQS}\;\;\lambda_{SSS}\big)\,
V_{QS}\,
\left(\begin{array}{c} 
\lambda_{QQQ}\\\lambda_{QSQ}\!=\!\lambda_{QQS}\\\lambda_{SSS} 
\end{array}\right)
\nonumber\\&
+\sum_{\substack{\cO_{\Delta,\ell}\in\,[N-1,1]\\\ell\,\text{even}}}
\big(\lambda_{QQ\cO}^{[N-1,1]_+}\;\;\lambda_{QS\cO}^{[N-1,1]_+}\big)\,
V_{4;\Delta,\ell}\,
\left(\begin{array}{c} \lambda_{QQ\cO}^{[N-1,1]_+}\\ \lambda_{QS\cO}^{[N-1,1]_+}\end{array}\right)
\nonumber\\&
+\sum_{\substack{\cO_{\Delta,\ell}\in\,[N-1,1]\\\ell\,\text{odd}}}
\big(\lambda_{QQ\cO}^{[N-1,1]_-}\;\;\lambda_{QS\cO}^{[N-1,1]_-}\big)\,
V_{5;\Delta,\ell}\,
\left(\begin{array}{c} \lambda_{QQ\cO}^{[N-1,1]_-}\\ \lambda_{QS\cO}^{[N-1,1]_-}\end{array}\right)
\nonumber\\&
+\sum_{\substack{\cO_{\Delta,\ell}\in\,\bullet\\\ell\,\text{even}}}
\big(\lambda_{QQ\cO}^{\bullet}\;\;\lambda_{SS\cO}^{\bullet}\big)\,
V_{6;\Delta,\ell}\,
\left(\begin{array}{c} \lambda_{QQ\cO}^{\bullet}\\ \lambda_{SS\cO}^{\bullet}\end{array}\right)
\nonumber\\&+\sum_{r=1,3} \sum_{\substack{\cO_{\Delta,\ell}\in\,\mathbf{r}\\\ell\,\text{even}}}
\bigl(\lambda_{QQ\cO}^{\mathbf{r}}\bigr)^2\,
W_{r;\Delta,\ell} + \sum_{\substack{\cO_{\Delta,\ell}\in\,[N-2,1,1]\\\ell\,\text{odd}}}
\bigl(\lambda_{QQ\cO}^{[N-2,1,1]}\bigr)^2\,
W_{2;\Delta,\ell}
\,.
\end{align}
where the representations in the last line are numbered as in Table~\ref{tab:frOPEsinglecorr}, namely $\mathbf{1} = [N-1,N-1,1,1]$ and $\mathbf{3} = [N-2,2]$. The matrices inside $V_{r;\Delta,\ell}$ are $2\times2$ while those in $V_{QS}$ are $3\times3$. The $W_{r;\Delta,\ell}$ are instead vectors. In order to simplify the notation we will abbreviate $F_{\pm,\Delta,\ell}^{ABCD}$ to $f_\pm$. The choices of the external operators are clear from the context. Indeed
\be
\begin{aligned}
&V_{QS} \supset \left(\begin{array}{ccc}
\# F_{\pm,\Delta_Q,0}^{QQQQ} &\# F_{\pm,\Delta_Q,0}^{QQQS}  &0\\
\# F_{\pm,\Delta_Q,0}^{QQQS} &\# F_{\pm,\Delta_S,0}^{QQQQ}\;\mathrm{or}\;\# F_{\pm,\Delta_Q,0}^{QSSQ} & \# F_{\pm,\Delta_S,0}^{SSQQ}\\
0&\# F_{\pm,\Delta_S,0}^{SSQQ}& \# F^{SSSS}_{\pm,\Delta_S,0}
\end{array}\right)\,,\\
&V_{6,\Delta,\ell} \supset \left(\begin{array}{cc}
\#\,F_{\pm,\Delta,\ell}^{QQQQ} &\#\,F_{\pm,\Delta,\ell}^{QQSS} \\
\#\,F_{\pm,\Delta,\ell}^{QQSS} &\#\,F_{\pm,\Delta,\ell}^{SSSS\phantom{^|}} \\
\end{array}\right)\,,\\
&V_{r,\Delta,\ell} \supset \left(\begin{array}{cc}
\#\,F_{\pm,\Delta,\ell}^{QQQQ} &\#\,F_{\pm,\Delta,\ell}^{QQQS} \\
\#\,F_{\pm,\Delta,\ell}^{QQQS} &\#\,F_{\pm,\Delta,\ell}^{QSSQ} \\
\end{array}\right)\,,\qquad &r=4,5\,,\\
&W_{r,\Delta,\ell} \supset \#\,F_{\pm,\Delta,\ell}^{QQQQ}\,,\qquad &r = 1,2,3\,.
\end{aligned}
\ee
Furthermore we will denote a sequence of $n$ zero matrices as $\mathbf{0}_n$. Finally, we also introduce some abbreviations for the various matrices that appear:
\be
\bfu = \left(\begin{matrix}1 &0\\0&0\end{matrix}\right)\,,\qquad
\bfl_\pm = \left(\begin{matrix}0&0\\0&f_\pm\end{matrix}\right)\,,\qquad
\bfs_\pm = \left(\begin{matrix}0&f_\pm\\f_\pm&0\end{matrix}\right)\,.
\ee
Now we are ready to show the explicit form of the crossing vectors and matrices

\makeatletter
\def\maketag@@@#1{\hbox{\m@th\normalfont\normalsize#1}}
\makeatother

\begin{footnotesize}

\begin{equation}
V_{4;\Delta,\ell} = \left(
\begin{aligned}
V_{[N-1,1]_+;\Delta,\ell}& \bfu\\
&\bfl_-\\&\bfl_+\\\tfrac12&\bfs_+\\&\mathbf0\\&\bfl_+\\&\mathbf0
\end{aligned}
\right)
\,,\qquad
V_{5;\Delta,\ell} = \left(
\begin{aligned}
V_{[N-1,1]_-;\Delta,\ell}& \bfu\\
&\bfl_-\\
&\bfl_+\\
&\mathbf0\\
\tfrac{1}{2} &\bfs_-\\
-&\bfl_+\\
&\mathbf0
\end{aligned}
\right)\,,\qquad
V_{6;\Delta,\ell} = \left(
\begin{aligned}
V_{\bullet;\Delta,\ell}& \bfu\\
-\tfrac{1}{2} &\bfs_-\\
\tfrac{1}{2} &\bfs_+\\
&\mathbf0\\
&\mathbf0\\
&\mathbf0\\
&\bfl_+
\end{aligned}
\right)\,.
\end{equation}

\end{footnotesize}

\begin{scriptsize}

\begin{equation}
V_{QS} = \left(
\begin{array}{c}
\left(
\begin{array}{ccc}
 0 & 0 & 0 \\
 0 & \frac{(N-1) (N+1) f_+}{N (N+2)} & 0 \\
 0 & 0 & 0 \\
\end{array}
\right)\\\left(
\begin{array}{ccc}
 0 & 0 & 0 \\
 0 & \frac{(N-1) (N+1) f_+}{2 (N-2) (N+2)} & 0 \\
 0 & 0 & 0 \\
\end{array}
\right)\\\left(
\begin{array}{ccc}
 0 & 0 & 0 \\
 0 & \frac{(N-1) (N+1) f_+}{4 (N-2) N} & 0 \\
 0 & 0 & 0 \\
\end{array}
\right)\\\left(
\begin{array}{ccc}
 f_+ & 0 & 0 \\
 0 & \frac{(N-1) N (N+1) f_+}{(N-2)^2 (N+2)^2} & 0 \\
 0 & 0 & 0 \\
\end{array}
\right)\\\left(
\begin{array}{ccc}
 -\frac{4 (N-2) (N+1) (N+2) f_-}{N^2 (N+3)} & 0 & 0 \\
 0 & -\frac{4 (N+1) f_-}{N (N+3)} & 0 \\
 0 & 0 & 0 \\
\end{array}
\right)\\\left(
\begin{array}{ccc}
 \frac{(N+2) f_-}{N} & 0 & 0 \\
 0 & -\frac{(N+1) f_-}{N+2} & 0 \\
 0 & 0 & 0 \\
\end{array}
\right)\\\left(
\begin{array}{ccc}
 0 & 0 & 0 \\
 0 & f_- & -\frac{f_-}{2} \\
 0 & -\frac{f_-}{2} & 0 \\
\end{array}
\right)\\\left(
\begin{array}{ccc}
 0 & 0 & 0 \\
 0 & f_+ & \frac{f_+}{2} \\
 0 & \frac{f_+}{2} & 0 \\
\end{array}
\right)\\\left(
\begin{array}{ccc}
 0 & \frac{f_+}{2} & 0 \\
 \frac{f_+}{2} & 0 & 0 \\
 0 & 0 & 0 \\
\end{array}
\right)\\\left(
\begin{array}{ccc}
 0 & 0 & 0 \\
 0 & 0 & 0 \\
 0 & 0 & 0 \\
\end{array}
\right)\\\left(
\begin{array}{ccc}
 0 & 0 & 0 \\
 0 & f_+ & 0 \\
 0 & 0 & 0 \\
\end{array}
\right)\\\left(
\begin{array}{ccc}
 0 & 0 & 0 \\
 0 & 0 & 0 \\
 0 & 0 & f_+ \\
\end{array}
\right)
\end{array}
\right)\,,\qquad
\begin{aligned}
&W_{1,\Delta,\ell} = \left(
\begin{aligned}
 V_{[N-1,N-1,1,1];\Delta,\ell} \;&\bfu \\ &\mathbf0_6
\end{aligned}
\right)\,,\\\\\\
&W_{2,\Delta,\ell} = \left(
\begin{aligned}
 V_{[N-2,1,1];\Delta,\ell} \;&\bfu \\ &\mathbf0_6
\end{aligned}
\right)\,,\\\\\\
&W_{3,\Delta,\ell} = \left(
\begin{aligned}
 V_{[N-2,2];\Delta,\ell} \;&\bfu \\ &\mathbf0_6
\end{aligned}
\right)\,.
\end{aligned}
\end{equation}

\end{scriptsize}

\noindent In the above equations $V_{\mathbf{irrep};\Delta,\ell}$ are the vectors appearing in \eqref{eq:CrossingEquationsSing}.

\section{Crossing equations for two adjoint scalars}

In this appendix we show the crossing equations for the mixed correlator system of two different adjoint scalars. We did not use these crossing equations in the body of the paper, but we wish to present them anyway as they could be useful for further generalizations of the present work. We use the notation and conventions set up in Section~\ref{sec:setup}.

\subsection[\texorpdfstring{The $Q\times Q'$ OPE}{The Q x Q' OPE}]{The $\boldsymbol{Q\times Q'}$ OPE}

We want to study the mixed system of adjoint representations. The most generic case consists in the OPE 
\be
Q(\mathbf{p}_1) \times Q'(\mathbf{p}_2) \,.
\ee
In the above product the representations exchanged are the same as in~\eqref{eq:exchangedIrreps} with the exception that the constraints on the spin parity are now modified. There are two major technical differences with the single correlator case
\begin{enumerate}
\item The complex representation $[N-2,1,1]$ and its conjugate $[N-1,N-1,2]$  contributed to the single correlator only through their sum. Here however both their sum and their difference appear.
\item Since all spins can be exchanged, now the two adjoints $[N-1,1]_\pm$ cannot be separated based on the parity of $\ell$ so there is a nontrivial (i.e. non diagonal) projector matrix from the space of three-point structures squared to the space of four-point structures.
\end{enumerate}

It will be important to study the conjugation properties of the various OPE coefficients to ensure that in the crossing equations we only have positive definite contributions. 
To do this one has to analyze the three-point functions. We will show explicitly the correlator of the adjoint since it is the only one having two structures. For the other representations we skip the analysis and report directly the results. The two structures for an exchanged adjoint read
\begin{equation}
\begin{aligned}
\langle Q(\mathbf{p}_1) Q'(\mathbf{p}_2) \mathcal{O}_{[N-1,1]}(\mathbf{p}_3)  \rangle&=\mathcal T_3\,\Bigl(\lambda_{QQ'\mathcal O}^{[N-1,1],+} \left((S_1\cdot \bar{S}_2)(S_2\cdot \bar{S}_3)(S_3\cdot \bar{S}_1) +(1\leftrightarrow2) \right) \\
 &+ i \lambda_{QQ'\mathcal O}^{[N-1,1],-} \left((S_1\cdot \bar{S}_2)(S_2\cdot \bar{S}_3)(S_3\cdot \bar{S}_1) - (1\leftrightarrow2) \right)\Bigr)\,,
\end{aligned}
\end{equation}
where $\mathcal{T}_3$ is the only three-point structure of two scalars and an operator of spin $\ell$
\begin{equation}
\mathcal{T}_3 = \frac{Z^{\mu_1}\cdots Z^{\mu_\ell} - \mathrm{traces}}{(x_{12}^2)^{\Delta_A+\Delta_B-\Delta_\cO + \ell}(x_{23}^2)^{\Delta_B+\Delta_\cO-\Delta_A - \ell}(x_{13}^2)^{\Delta_\cO+\Delta_A-\Delta_B - \ell}} \,,\qquad Z^\mu \equiv \frac{x_{13}^\mu}{x_{13}^2} - \frac{x_{23}^\mu}{x_{23}^2}\,.
\end{equation}
Notice that the two tensor structures have different transformation properties under conjugation and permutation. Also notice the factor of $i$ in the second structure. The symmetry and conjugations properties therefore read
\be
\lambda_{QQ'\mathcal O}^{[N-1,1],\pm}=  \bigl(\lambda_{QQ'\mathcal O}^{[N-1,1],\pm}\bigr)^*\,, \qquad  \lambda_{Q'Q\mathcal O}^{[N-1,1],\pm}  = \pm(-1)^{\ell_{\mathcal{O}}}  \lambda_{QQ'\mathcal O}^{[N-1,1],\pm} \label{adjointlambda}\,.
\ee
Now let us proceed with the remaining representations. For $[N-1,N-1,1,1]$ we have 
\be
\lambda_{QQ'\mathcal O}^{[N-1,N-1,1,1]}=  \bigl(\lambda_{QQ'\mathcal O}^{[N-1,N-1,1,1]}\bigr)^*\,, \qquad  \lambda_{Q'Q\mathcal O}^{[N-1,N-1,1,1]}  = (-1)^{\ell_{\mathcal{O}}}  \lambda_{QQ'\mathcal O}^{[N-1,N-1,1,1]}\,.
\ee
Similar properties hold for the singlet and the $[N-2,2]$

\begin{align}
\lambda_{QQ'\mathcal O}^{\bullet} &=( \lambda_{QQ'\mathcal O}^{\bullet})^*\,, \quad  &\lambda_{Q'Q\mathcal O}^{\bullet}  &= (-1)^{\ell_{\mathcal{O}}}  \lambda_{QQ'\mathcal O}^{\bullet} \,.
\\
\lambda_{QQ'\mathcal O}^{[N-2,2]} &=  \bigl(\lambda_{QQ'\mathcal O}^{[N-2,2]}\bigr)^*\,, \quad  &\lambda_{Q'Q\mathcal O}^{[N-2,2]}  &= (-1)^{\ell_{\mathcal{O}}}  \lambda_{QQ'\mathcal O}^{[N-2,2]}\,.
\end{align}
Finally, the $[N-1,N-1,2]$ is the only one that has a nontrivial conjugation property
\begin{subequations}
\begin{align}
\lambda_{QQ'\mathcal O}^{[N-1,N-1,2]}&=  \bigl(\lambda_{QQ'\mathcal O}^{[N-2,1,1]}\bigr)^*\,, \label{conjugation}\\
  \lambda_{Q'Q\mathcal O}^{[N-1,N-1,2]}  &= (-1)^{\ell_{\mathcal{O}}+1}  \lambda_{QQ'\mathcal O}^{[N-1,N-1,2]}\,, \label{complexlambda1}\\
  \lambda_{Q'Q\mathcal O}^{[N-2,1,1]}  &= (-1)^{\ell_{\mathcal{O}}+1}  \lambda_{QQ'\mathcal O}^{[N-2,1,1]}\,.  \label{complexlambda2}
\end{align}
\end{subequations}

\subsection{Four-point structures}

We will now write down the possible four-point tensor structures for any combination of the operators $Q$ and $Q'$. The four-point structure $\mathcal{T}_4$ is defined in \eqref{eq:T4struct}, with $\Delta_A \equiv \Delta_{Q^A}$. Since we assume a $\Z_2$ symmetry that acts trivially on $Q$ and flips the sign of $Q'$, the only four-point functions that we have to consider have an even number of $Q'$ operators. Nevertheless, let us for brevity write down a general formula (with $Q^1\equiv Q$ and $Q^2\equiv Q'$)
\be
\langle Q^A(\mathbf{p}_1) Q^B(\mathbf{p}_2) Q^C(\mathbf{p}_3) Q^D(\mathbf{p}_4)\rangle = \cT_4\,\sum_{r=1}^7 \mathfrak{Re}\;T^{(r)}\,f^{ABCD}_r(u,v)\,,
\ee
The structures $T^{(1)}$ through $T^{(6)}$ are even or odd under $1\leftrightarrow 2$, while $T^{(7)}$ does not transform in a definite way. The real part will be relevant only for $T^{(2)}$, all other structures are taken to be real. Let us define as a shorthand
\be
\mathcal{S}_{ijkl} = (S_1\cdot \bar{S}_i)(S_2\cdot \bar{S}_j)(S_3\cdot \bar{S}_k)(S_4\cdot \bar{S}_l)\,.
\ee
The counting of all possible $\mathcal{S}_{ijkl}$ is simple: due to $S_i\cdot\bar{S}_i=0$ we have to count all permutation of four elements without 1-cycles, which results nine. However two structures can be combined in the complex combination $T^{(2)}$ and one structure\footnote{namely $T^{(8)}=\mathcal{S}_{2413}-\mathcal{S}_{3142}$.\label{footnoteT8}} never appears because we only consider correlators with an even number of $Q'$. The most convenient basis for the $T^{(r)}$ is the one where $f_r$ gets contributions from a single OPE channel. The choice realizing this requirement is
\be
\begin{aligned}
T^{(1)} &= \frac{\mathcal{S}_{(43)(21)}}4 - \frac{\mathcal{S}_{2341}+\mathcal{S}_{2413}+\mathcal{S}_{4123}+\mathcal{S}_{3142}}{4(N+2)}+\frac{\mathcal{S}_{2143}}{2(N+1)(N+2)}\,,\\
T^{(2)} &= \frac{\mathcal{S}_{4321}-\mathcal{S}_{3412}}{2} - \frac{\mathcal{S}_{2341}-\mathcal{S}_{2413}+\mathcal{S}_{4123}-\mathcal{S}_{3142}}{2N}+\frac{\mathcal{S}_{3421}-\mathcal{S}_{4312}}{2}\,,\\
\bar{T}^{(2)} &= \frac{\mathcal{S}_{4321}-\mathcal{S}_{3412}}{2} - \frac{\mathcal{S}_{2341}-\mathcal{S}_{2413}+\mathcal{S}_{4123}-\mathcal{S}_{3142}}{2N}-\frac{\mathcal{S}_{3421}-\mathcal{S}_{4312}}{2}\,,\\
T^{(3)} &=\frac{\mathcal{S}_{[43][21]}}{4}  - \frac{\mathcal{S}_{2341}+\mathcal{S}_{2413}+\mathcal{S}_{4123}+\mathcal{S}_{3142}}{4(N-2)}+\frac{\mathcal{S}_{2143}}{2(N-1)(N-2)}\,\\
T^{(4)} &=\mathcal{S}_{2341}+\mathcal{S}_{2413}+\mathcal{S}_{4123}+\mathcal{S}_{3142} - \frac{4\,\mathcal{S}_{2143}}{N}\,,\\
T^{(5)} &= -\mathcal{S}_{2341}+\mathcal{S}_{2413}-\mathcal{S}_{4123}+\mathcal{S}_{3142}\,,\\
T^{(6)} &= \mathcal{S}_{2143}\,,\\
T^{(7)} &= 2i\,(\mathcal{S}_{2341}-\mathcal{S}_{4123})\,.\\
\end{aligned}
\ee
As mentioned before, the structure $T^{(2)}$ is complex.\footnote{These structures are identical to those of the single correlator analysis~\eqref{eq:structsSingle} with the exception that $T^{(2)}$ there is $\frac12\bigl(T^{(2)} + \bar{T}^{(2)}\bigr)$ here, i.e. the real part.} This is because it contains the contributions from operators in the complex representation $[N-2,1,1]$. Notice however that, thanks to the symmetry property \eqref{conjugation}, if $A=C$ and $B=D$ then $f_2$ is real and only $\mathfrak{Re}\,T^{(2)}$ survives. A clarification for the adjoint representation is in order. Since the three-point functions with three adjoints have two structures we would expect an OPE contribution of the form
\be
\cT_4^{-1}\langle Q^A Q^B Q^C Q^D\rangle \supset T^{(r)}\,\lambda_{Q^AQ^B\cO}^{[N-1,1]_a} \lambda_{Q^CQ^D\cO}^{[N-1,1]_b}\,\Pi_{ab;r}G_{\Delta,\ell}(u,v)\,,\qquad a,b = \pm\,.
\ee
where $\Pi_{ab;r}$ is a projector relating the product of three-point tensor structures to four-point tensor structures. The mixed terms $a\neq b$ can only appear when either $A\neq B$ or $C\neq D$ due to the spin parity being opposite. But since we only consider the cases with an even number of $Q'$, this only occurs for $\langle QQ'QQ'\rangle$ (and $(12)$ or $(34)$ exchanges thereof). Thus there are only these combinations
\be
\Pi_{++;r} \propto \delta_{r4}\,,\qquad \Pi_{--;r} \propto \delta_{r5}\,,\qquad \mbox{and}\qquad (\Pi_{+-;r} + \Pi_{-+;r}) \propto \delta_{r7}\,.
\ee
In the more general case without $\Z_2$ symmetry, $\Pi_{\pm\mp;k}$ would appear separately and this would require another structure $T^{(8)}$ (see footnote \ref{footnoteT8}).
\par
The OPE channels associated to each partial wave $f_r^{ABCD}$ are summarized in \tablename~\ref{tab:frOPE}. Let us write down their explicit definition. For brevity we denote $\delta \equiv \Delta_Q - \Delta_{Q'}$. For $r=1,3,6$ one has
\be
\begin{aligned}
f_r^{1122}(u,v) &= \sum_{\substack{\cO_{\Delta,\ell} \in\, \mathbf{r}\\\ell\;\text{even}}}\,\lambda_{QQ\cO}^{\mathbf{r}} \,\lambda_{Q'Q'\cO}^{\mathbf{r}} \,g^{0,0}_{\Delta,\ell}(u,v)\,,\\
f_r^{1212}(u,v) &= \sum_{\cO_{\Delta,\ell} \in \,\mathbf{r}}\,(-1)^{\ell}\big(\lambda_{QQ'\cO}^{\mathbf{r}} \big)^2\,g^{\delta,\delta}_{\Delta,\ell}(u,v)\,,\\
\end{aligned}
\ee
with $\mathbf{1}=[N-1,N-1,1,1]$, $\mathbf{3}=[N-2,2]$ and $\mathbf{6}=\bullet$. The other representations follow:

\be
\begin{aligned}
f_2^{1122}(u,v) &= \sum_{\substack{\cO_{\Delta,\ell} \in [N-2,1,1]\\\ell\;\text{odd}}}\,\lambda_{QQ\cO}^{[N-2,1,1]} \lambda_{Q'Q'\cO}^{[N-1,N-1,2]} \,g^{0,0}_{\Delta,\ell}(u,v)\,,\\
f_2^{1212}(u,v) &= \sum_{\cO_{\Delta,\ell} \in [N-2,1,1]}\,(-1)^{\ell}\,\lambda_{QQ'\cO}^{[N-2,1,1]} \lambda_{QQ'\cO}^{[N-1,N-1,2]}\,g^{\delta,\delta}_{\Delta,\ell}(u,v)\,,\\
\end{aligned}
\ee

\be
\begin{aligned}
f_4^{1122}(u,v) &= \sum_{\substack{\cO_{\Delta,\ell} \in [N-1,1]\\\ell\;\text{even}}}\,\lambda_{QQ\cO}^{[N-1,1]_+} \lambda_{Q'Q'\cO}^{[N-1,1]_+} \,g^{0,0}_{\Delta,\ell}(u,v)\,,\\
f_4^{1212}(u,v) &= \sum_{\cO_{\Delta,\ell} \in [N-1,1]}\,(-1)^{\ell}\,\big(\lambda_{QQ'\cO}^{[N-1,1]_+} \big)^2\,g^{\delta,\delta}_{\Delta,\ell}(u,v)\,,\\
\end{aligned}
\ee
\be
\begin{aligned}
f_5^{1122}(u,v) &= \sum_{\substack{\cO_{\Delta,\ell} \in [N-1,1]\\\ell\;\text{odd}}}\,\lambda_{QQ\cO}^{[N-1,1]_-} \lambda_{Q'Q'\cO}^{[N-1,1]_-} \,g^{0,0}_{\Delta,\ell}(u,v)\,,\\
f_5^{1212}(u,v) &= \sum_{\cO_{\Delta,\ell} \in [N-1,1]}\,(-1)^{\ell}\,\big(\lambda_{QQ'\cO}^{[N-1,1]_-} \big)^2\,g^{\delta,\delta}_{\Delta,\ell}(u,v)\,,\\
\end{aligned}
\ee
\be
\begin{aligned}
f_7^{1212}(u,v) &= \sum_{\cO_{\Delta,\ell} \in [N-1,1]}\,(-1)^{\ell}\,\lambda_{QQ'\cO}^{[N-1,1]_+}\,\lambda_{QQ'\cO}^{[N-1,1]_-}\,g^{\delta,\delta}_{\Delta,\ell}(u,v)\,.\\
\end{aligned}
\ee
The partial waves $f_r^{1221}$ are obtained by removing the $(-1)^\ell$ from $f_r^{1212}$, except when $r=2,5$ where we have to further add a minus sign due to \eqref{adjointlambda}, \eqref{complexlambda1} and \eqref{complexlambda2}.
The complex conjugate of $f_2$ is obtained using \eqref{conjugation} on the OPE coefficients. The conformal blocks are normalized according to the convention expressed in~\eqref{eq:cbconvention}. That normalization implies $g_{\Delta,\ell}^{\delta,-\delta}\big(\frac14,\frac14\big)\geq0$ for all $\Delta,\ell,\delta$.
\begin{table}
\centering
\begin{tabular}{lllll}
\toprule
$f_r^{ABCD}$ & $\langle Q^A Q^B \cO|\phantom{\displaystyle\widehat{O}}$ &  $ |\bar{\cO} Q^C Q^D \rangle$ & If $A = B$ & If $A=C$\\
\midrule
$f_1$ & $[N-1,N-1,1,1]$ & $[N-1,N-1,1,1]$  & only $\ell$ even \\
$f_2$ & $[N-2,1,1]$ & $[N-1,N-1,2]$ & only $\ell$ odd & $f_2 = \bar{f}_2$\\
$\bar{f}_2$ & $[N-1,N-1,2]$ & $[N-2,1,1]$ & only $\ell$ odd & $f_2 = \bar{f}_2$\\
$f_3$ & $[N-2,2]$ & $[N-2,2]$ & only $\ell$ even\\
$f_4$ & $[N-1,1]_+$  & $[N-1,1]_+$ & only $\ell$ even\\
$f_5$ & $[N-1,1]_-$ & $[N-1,1]_-$ & only $\ell$ odd\\
$f_6$ & $\bullet$ & $\bullet$ & only $\ell$ even\\
$f_7$ & $[N-1,1]_\pm$ & $[N-1,1]_\mp$ & not present & only if $A \neq B$ \\
\bottomrule
\end{tabular}
\caption{Operator contributions to the partial wave $f_r^{ABCD}$.}\label{tab:frOPE}
\end{table}

\subsection{Crossing equations}
Using the definition of $F_{\pm,\Delta,\ell}$ in \eqref{eq:FHdefMixed}, the crossing equations take the form
\begin{align}
(1\;\;1)\, V_{6;\lsp0,0}\left(\begin{matrix}1\\1\end{matrix}\right)=&\sum_{r=1,3,6} \sum_{\substack{\cO_{\Delta,\ell}\in\,\mathbf{r}\\\ell\,\text{even}}}
\big(\lambda_{QQ\cO}^{\mathbf{r}}\;\;\lambda_{Q'Q'\cO}^{\mathbf{r}}\big)\,
V_{r;\Delta,\ell}\,
\left(\begin{array}{c} \lambda_{QQ\cO}^{\mathbf{r}}\\ \lambda_{Q'Q'\cO}^{\mathbf{r}}\end{array}\right)
\nonumber\\&
+\sum_{\substack{\cO_{\Delta,\ell}\in\,[N-2,1,1]\\\ell\,\text{odd}}}
\big(\lambda_{QQ\cO}^{\mathfrak{Re}}\;\;\lambda_{QQ\cO}^{\mathfrak{Im}}\;\;\lambda_{Q'Q'\cO}^{\mathfrak{Re}} \;\;\lambda_{Q'Q'\cO}^{\mathfrak{Im}} \big)\,
V_{2;\Delta,\ell}\,
\left(\begin{array}{c} 
\lambda_{QQ\cO}^{\mathfrak{Re}\phantom{^|}}\\\lambda_{QQ\cO}^{\mathfrak{Im}\phantom{^|}}\\\lambda_{Q'Q'\cO}^{\mathfrak{Re}\phantom{^|}} \\\lambda_{Q'Q'\cO}^{\mathfrak{Im}\phantom{^|}} 
\end{array}\right)
\nonumber\\&
+\sum_{\substack{\cO_{\Delta,\ell}\in\,[N-1,1]\\\ell\,\text{even}}}
\big(\lambda_{QQ\cO}^{[N-1,1]_+}\;\;\lambda_{Q'Q'\cO}^{[N-1,1]_+}\big)\,
V_{4;\Delta,\ell}\,
\left(\begin{array}{c} \lambda_{QQ\cO}^{[N-1,1]_+}\\ \lambda_{Q'Q'\cO}^{[N-1,1]_+}\end{array}\right)
\nonumber\\&
+\sum_{\substack{\cO_{\Delta,\ell}\in\,[N-1,1]\\\ell\,\text{odd}}}
\big(\lambda_{QQ\cO}^{[N-1,1]_-}\;\;\lambda_{Q'Q'\cO}^{[N-1,1]_-}\big)\,
V_{5;\Delta,\ell}\,
\left(\begin{array}{c} \lambda_{QQ\cO}^{[N-1,1]_-}\\ \lambda_{Q'Q'\cO}^{[N-1,1]_-}\end{array}\right)\,.
\end{align}
\be
\begin{aligned}
0=&\sum_{r=1,3,6} \sum_{\cO_{\Delta,\ell}\in\,\mathbf{r}}(-1)^\ell\big(\lambda_{QQ'\cO}^{\mathbf{r}}\big)^2\,W_{r;\Delta,\ell}
\\&
+\sum_{\cO_{\Delta,\ell}\in\,[N-2,1,1]}(-1)^\ell\big|\lambda_{QQ'\cO}^{[N-2,1,1]}\big|^2\,W_{2;\Delta,\ell}
\\&
+\sum_{\cO_{\Delta,\ell}\in\,[N-1,1]}(-1)^\ell\;\big(\lambda_{QQ'\cO}^{[N-1,1]_+}\;\;\lambda_{QQ'\cO}^{[N-1,1]_-}\big)\,
W_{\mathbf{adj};\Delta,\ell}\,\left(\begin{array}{c} \lambda_{QQ'\cO}^{[N-1,1]_+}\phantom{\Big|}\\\lambda_{QQ'\cO}^{[N-1,1]_-}\end{array}\right)\,.
\end{aligned}
\ee
where $V_{r\neq2;\Delta,\ell}$ and $W_{\mathbf{adj};\Delta,\ell}$ are $35$ component vectors of $2\times2$ matrices, while $V_{2;\Delta,\ell}$ is a $35$ component vector of $4\times4$ matrices and the remaining $W_{r;\Delta,\ell}$ are ordinary $35$ component vectors. We also defined
\be
\lambda_{QQ\cO}^{\mathfrak{Re}} \equiv \mathfrak{Re}\,\lambda_{QQ\cO}^{[N-2,1,1]}\,,\qquad
\lambda_{QQ\cO}^{\mathfrak{Im}} \equiv \mathfrak{Im}\,\lambda_{QQ\cO}^{[N-2,1,1]}\,.
\ee
\par
In order to make the formulas more compact we abbreviate $F_{\pm,\Delta,\ell}^{ABCD}$ with $f_\pm$. The indices $ABCD$ can be deduced from context, namely the $2\times2$ matrices are of the form
\be
V_{r;\Delta,\ell} \supset \left(\begin{array}{cc}
\#\,F_{\pm,\Delta,\ell}^{1111} &\#\,F_{\pm,\Delta,\ell}^{2211} \\
\#\,F_{\pm,\Delta,\ell}^{2211} &\#\,F_{\pm,\Delta,\ell}^{2222\phantom{^|}} \\
\end{array}\right)\,,\qquad
W_{\mathbf{adj};\Delta,\ell} \supset \left(\begin{array}{cc}
\#\,F_{\pm,\Delta,\ell}^{1212} &\#\,F_{\pm,\Delta,\ell}^{1221} \\
\#\,F_{\pm,\Delta,\ell}^{1221} &\#\,F_{\pm,\Delta,\ell}^{1212\phantom{^|}} \\
\end{array}\right)\,.
\ee
The $4\times4$ is similar to $V_{r;\Delta,\ell}$ when split in $2\times2$ blocks
\be
V_{2;\Delta,\ell} \supset \left(\begin{array}{c|c}
\#\,F_{\pm,\Delta,\ell}^{1111} &\#\,F_{\pm,\Delta,\ell}^{2211} \\
\hline
\#\,F_{\pm,\Delta,\ell}^{2211} &\#\,F_{\pm,\Delta,\ell}^{2222\phantom{^|}} \\
\end{array}\right)\,,
\ee
and finally $W_{r,\Delta,\ell}$ has $F^{1212}_{\pm,\Delta,\ell}$ in the last $7$ entries and $F^{1221}_{\pm,\Delta,\ell}$ in the remaining ones. Furthermore we will introduce some abbreviations for the various matrices that appear:
\be
\bfu_\pm = \left(\begin{matrix}f_\pm &0\\0&0\end{matrix}\right)\,,\qquad
\bfl_\pm = \left(\begin{matrix}0&0\\0&f_\pm\end{matrix}\right)\,,\qquad
\bfs_\pm = \left(\begin{matrix}0&f_\pm\\f_\pm&0\end{matrix}\right)\,,\qquad
\bfa_\pm = \left(\begin{matrix}0&-f_\pm\\f_\pm&0\end{matrix}\right)\,.\label{matrixnotation}
\ee
We use $\otimes$ to denote the Kronecker product.\footnote{
e.g. \[
\left(\begin{matrix}a&b\\c&d\end{matrix}\right)\otimes \mathds{1} = \left(
\begin{array}{cccc}
 a & 0 & b & 0 \\
 0 & a & 0 & b \\
 c & 0 & d & 0 \\
 0 & c & 0 & d \\
\end{array}
\right)\,.
\]
}
Finally we indicate with $0_n$ a string of $n$ zeros and with $\mathbf{0}_n$ a string of $n$ zero matrices.

\makeatletter
\def\maketag@@@#1{\hbox{\m@th\normalfont\normalsize#1}}
\makeatother

\begingroup
\addtolength{\jot}{-.3em}
\be\scriptsize
V_{1;\Delta,\ell} = \left(
\begin{aligned}
&\bfu_- \\
&\bfzr_3 \\
&\bfu_+ \\
&\bfzr\\
\tfrac{N^2+N+2}{8(N+1)(N+2)}&\bfs_- \\
\tfrac{N(N+3)}{8(N+1)(N+2)}&\bfs_- \\
\tfrac{N+3}{8(N+1)}&\bfs_- \\
\tfrac{N^3 (N+3)}{16 (N-2) (N+1) (N+2)^2} & \bfs_- \\
-\tfrac{N+3}{16 (N+1)} &\bfs_- \\
\tfrac{N^2 (N+3)}{8 (N-1) (N+1)^2} & \bfs_- \\
&\bfzr \\
-\tfrac{N^2+N+2}{8 (N+1) (N+2)} & \bfs_+ \\
-\tfrac{N (N+3)}{8 (N+1) (N+2)} & \bfs_+ \\
-\tfrac{N+3}{8 (N+1)} & \bfs_+ \\
\tfrac{-N^3 (N+3)}{16 (N-2) (N+1) (N+2)^2} & \bfs_+ \\
\tfrac{N+3}{16 (N+1)} & \bfs_+ \\
-\tfrac{N^2 (N+3)}{8 (N-1) (N+1)^2} & \bfs_+\\
&\bfzr_3 \\
&\bfl_- \\
&\bfzr_3 \\
&\bfl_+ \\
&\bfzr_8\\
\end{aligned}
\right)\,,\qquad
V_{2;\Delta,\ell} = \left(
\begin{aligned}
&\bfzr \\
\bfu_-&\otimes \unit \\
&\bfzr_3 \\
\bfu_+&\otimes \unit \\
\tfrac{N-2}{2N}\,\bfs_-&\otimes \unit \\
\tfrac12\,\bfs_-&\otimes \unit\\
\tfrac{N+2}{2N}\,\bfs_-&\otimes \unit \\
\tfrac{N}{2 (N^2-4)}\,\bfs_-&\otimes \unit \\
\bfzr \\
-\tfrac{N^2-4}{2(N^2-1)}\,\bfs_-&\otimes \unit \\
\bfzr \\
-\tfrac{N-2}{2 N}\,\bfs_+&\otimes \unit \\
\tfrac12\,\bfs_+&\otimes \unit \\
-\tfrac{N+2}{2 N}\,\bfs_+&\otimes \unit \\
-\tfrac{N}{2 (N^2-4)}\bfs_+&\otimes \unit \\
\bfzr \\
\tfrac{N^2-4}{2(N^2-1)}\,\bfs_+&\otimes \unit \\
\bfzr \\
\bfa_-&\otimes\Big(\scriptsize\begin{matrix}0&1\\-1&0\end{matrix}\Big)\\
\bfa_+&\otimes\Big(\scriptsize\begin{matrix}0&1\\-1&0\end{matrix}\Big)\\
&\bfzr \\
\bfl_-&\otimes \unit \\
&\bfzr_3 \\
\bfl_+&\otimes \unit \\
&\bfzr_7
\end{aligned}
\right)\,,\qquad
\ee

\be\scriptsize
V_{3;\Delta,\ell} = \left(
\begin{aligned}
&\bfzr_2 \\
&\bfu_-&\\
&\bfzr \\
-\tfrac{(N-3) (N+1)}{(N-1) (N+3)}&\bfu_+&\\
\tfrac{N^2(N-3)}{(N^2-4) (N-1)} & \bfu_+ \\
\tfrac{N-3}{8 (N-1)} & \bfs_- \\
\tfrac{(N-3) N}{8 (N-2) (N-1)}& \bfs_-\\
\tfrac{N^2-N+2}{8 (N-2) (N-1)}& \bfs_-\\
\tfrac{-N^3(N-3)}{16 (N-2)^2 (N-1) (N+2)}& \bfs_-\\
\tfrac{N-3}{16 (N-1)}& \bfs_-\\
\tfrac{(N-3) N^2}{8 (N-1)^2 (N+1)}& \bfs_- \\
&\bfzr \\
-\tfrac{N-3}{8 (N-1)} & \bfs_+ \\
-\tfrac{(N-3) N}{8 (N-2) (N-1)} & \bfs_+ \\
-\tfrac{N^2-N+2}{8 (N-2) (N-1)} & \bfs_+ \\
\tfrac{(N-3) N^3}{16 (N-2)^2 (N-1) (N+2)} & \bfs_+ \\
-\tfrac{N-3}{16 (N-1)} & \bfs_+ \\
-\tfrac{N^2(N-3) }{8 (N-1)^2 (N+1)} & \bfs_+ \\
&\bfzr_5  \\
&\bfl_-&\\
&\bfzr \\
-\tfrac{(N-3) (N+1)}{(N-1) (N+3)}&\bfl_+&\\
\tfrac{N^2(N-3)}{(N^2-4) (N-1)} & \bfl_+ \\
&\bfzr_7
\end{aligned}
\right)\,,\qquad
V_{4;\Delta,\ell} = \left(
\begin{aligned}
&\bfzr_3 \\
-\tfrac{4 (N+1) (N^2-4)}{N^2 (N+3)}&\bfu_-\\
\tfrac{2 (N+2)}{N}&\bfu_+\\
&\bfu_+\\
\tfrac{N-2}{N}&\bfs_-\\
\tfrac{2}{N}&\bfs_-\\
-\tfrac{N+2}{N}&\bfs_-\\
\tfrac{N^2-12}{4 (N^2-4)}&\bfs_-\\
\tfrac{N^2-4}{4 N^2}&\bfs_-\\
\tfrac{N^2-4}{ N (N^2-1)}&\bfs_-\\
&\bfzr \\
-\tfrac{N-2}{N}&\bfs_+\\
-\tfrac2N &\bfs_+\\
\tfrac{N+2}{N}&\bfs_+\\
-\tfrac{N^2-12}{4 (N^2-4)}&\bfs_+\\
-\tfrac{N^2-4}{4 N^2}&\bfs_+\\
-\tfrac{N^2-4}{N (N^2-1)}&\bfs_+\\
&\bfzr_6 \\
&\bfl_-\\
-\tfrac{4 (N+1) (N^2-4)}{N^2 (N+3)}&\bfl_+\\
\tfrac{2 (N+2)}{N}&\bfl_+\\
&\bfzr_7
\end{aligned}
\right)\,.
\ee

\be\scriptsize
V_{5;\Delta,\ell} = \left(
\begin{aligned}
-\tfrac{4 (N+1)}{N+2}&\bfu_- \\
\tfrac{4 N}{N^2-4}&\bfu_- \\
\tfrac{4 (N-1)}{N-2}&\bfu_- \\
\tfrac{N^4}{(N^2-4)^2}&\bfu_- \\
\tfrac{4 (N+1)}{N+3}&\bfu_+ \\
-\tfrac{2 N}{N+2}&\bfu_+ \\
-&\bfs_- \\
&\bfzr \\
&\bfs_- \\
\tfrac{N^2}{4 (N^2-4)}&\bfs_- \\
\tfrac14 &\bfs_- \\  
\tfrac{N}{N^2-1}&\bfs_- \\
&\bfzr \\
&\bfs_+ \\
&\bfzr \\
-&\bfs_+ \\
-\tfrac{N^2}{4 (N^2-4)}&\bfs_+ \\
-\tfrac14 &\bfs_+ \\
-\tfrac{N}{N^2-1}&\bfs_+ \\
&\bfzr_3 \\
-\tfrac{4 (N+1)}{N+2}&\bfl_- \\
\tfrac{4 N}{N^2-4}&\bfl_- \\
\tfrac{4 (N-1)}{N-2}&\bfl_- \\
\tfrac{N^4}{(N^2-4)^2}&\bfl_- \\
\tfrac{4 (N+1)}{N+3}&\bfl_+ \\
-\tfrac{2 N}{N+2}&\bfl_+ \\
&\bfzr_7
\end{aligned}
\right)\,,\qquad
V_{6;\Delta,\ell} = \left(
\begin{aligned}
\tfrac{N^2-1}{N (N+2)} &\bfu_- \\
\tfrac{N^2-1}{N^2-4} &\bfu_- \\
\tfrac{N^2-1}{N (N-2)} &\bfu_- \\
\tfrac{N (N^2-1)}{(N^2-4)^2} &\bfu_- \\
-\tfrac{4 (N+1)}{N (N+3)} &\bfu_+ \\
-\tfrac{2 (N+1)}{N+2} &\bfu_+ \\
\tfrac12 &\bfs_- \\
-\tfrac12 &\bfs_- \\
\tfrac12 &\bfs_- \\
\tfrac{N}{4 (N^2-4)} &\bfs_- \\
\tfrac{1}{4N} &\bfs_- \\
\tfrac{1}{2 (N^2-1)} &\bfs_- \\
&\bfzr \\
-\tfrac12 &\bfs_+ \\
\tfrac12 &\bfs_+ \\
- \tfrac12 &\bfs_+ \\
-\tfrac{N}{4 (N^2-4)} &\bfs_+ \\
-\tfrac{1}{4 N}&\bfs_+ \\
-\tfrac{1}{2 (N^2-1)} &\bfs_+ \\
&\bfzr_3 \\
\tfrac{N^2-1}{N (N+2)} &\bfl_- \\
\tfrac{N^2-1}{N^2-4} &\bfl_- \\
\tfrac{N^2-1}{N (N-2)} &\bfl_- \\
\tfrac{N (N^2-1)}{(N^2-4)^2} &\bfl_- \\
-\tfrac{4 (N+1)}{N (N+3)} &\bfl_+ \\
-\tfrac{2 (N+1)}{N+2} &\bfl_+ \\
&\bfzr_7
\end{aligned}
\right)\,.
\ee

\be\scriptsize
W_{1;\Delta,\ell} = \left(
\begin{aligned}
&0_6\\
&f_- \\
&0_6\\
&f_+ \\
&0_{14}\\
& f_- \\
&0_3\\
&f_+ \\
&0_2\\
\end{aligned}
\right)\,,\qquad
W_{2;\Delta,\ell} = \left(
\begin{aligned}
&0_7 \\
2&f_- \\
&0_6 \\
2&f_+ \\
&0_{14} \\
2&f_- \\
&0_3 \\
2&f_+ \\
&0 \\
\end{aligned}
\right)\,,\qquad
W_{3;\Delta,\ell} = \left(
\begin{aligned}
& 0_8\\
&f_- \\
& 0_6\\
& f_+ \\
& 0_{14}\\
& f_- \\
& 0\\
-\tfrac{(N-3) (N+1)}{(N-1) (N+3)} & f_+ \\
\tfrac{N^2 (N-3)}{(N-1) (N^2-4)} & f_+ \\
& 0\\
\end{aligned}
\right)\,.
\ee

\be\scriptsize
W_{\mathbf{adj};\Delta,\ell} = \left(
\begin{aligned}
& \bfzr_9 \\
& \bfu_-\\
&\bfl_- \\
& \bfzr \\
\tfrac12 & \bfs_- \\
& \bfzr_3 \\
& \bfu_+\\
&\bfl_+ \\
& \bfzr \\
\tfrac12 &\bfs_+ \\
& \bfzr_8 \\
-\tfrac{4 (N+1)}{N+2} & \bfl_- \\
\tfrac{4 N}{N^2-4} & \bfl_- \\
\tfrac{4 (N-1)}{N-2} & \bfl_- \\
\tfrac{1}{(N^2-4)^2} &\Big(\scriptsize\begin{matrix}(N^2-4)^2 f_- & 0\\0&N^4 f_-\end{matrix}\Big)\\
\tfrac{4 (N+1)}{N^2 (N+3)} &  \Big(\scriptsize\begin{matrix}-(N^2-4) f_+ & 0\\0&N^2 f_+\end{matrix}\Big)\\
\tfrac{2}{N (N+2)} & \Big(\scriptsize\begin{matrix}(N+2)^2 f_- & 0\\0&-N^2 f_-\end{matrix}\Big)\\
\tfrac12 & \bfs_+
\end{aligned}
\right)\,,\qquad
W_{6;\Delta,\ell} = \left(
\begin{aligned}
& 0_{11} \\
&f_- \\
& 0_6 \\
&f_+ \\
& 0_9 \\
\tfrac{N^2-1}{N (N+2)} & f_- \\
\tfrac{N^2-1}{N^2-4} & f_- \\
\tfrac{N^2-1}{(N-2) N} & f_- \\
\tfrac{N (N^2-1)}{(N^2-4)^2} & f_- \\
-\tfrac{4 (N+1)}{N (N+3)} & f_+ \\
-\tfrac{2 (N+1)}{N+2} & f_+ \\
& 0
\end{aligned}
\right)\,.
\ee
\endgroup

\section{Unitary versus orthogonal group bootstrap}
\label{app:RelationshipToVectorBootstrap}

In this appendix we prove a correspondence between bounds obtained bootstrapping $SU(N)$ adjoint scalars and bounds obtained bootstrapping $O(N')$ fundamental scalars, where $N' = N^2-1$. In particular, we prove that the upper bounds on the dimension of the lightest singlet are identical. Similar theorems hold for other pairs of representations and groups, such as fundamentals in $O(N(N+1)/2-1)$ vs. rank-2 tensors in $O(N)$~\cite{Reehorst:2020phk} and bifundamentals in $SU(N)\times SU(N)$ vs. fundamentals in $O(2N)$~\cite{Li:2020bnb}. The proof that we will present here follows exactly the same ideas.

As a first trivial observation we note that, when $N' = N^2-1$, the fundamental of $O(N')$ decomposes in the adjoint of $SU(N)$. Indeed their dimensions agree. The OPE of two $O(N')$ fundamentals contains the singlet ($S$), the rank-2 tensor ($T$) and the antisymmetric ($A$), while the OPE of two adjoints in $SU(N)$ contains the six representations shown in~\eqref{eq:exchangedIrreps}. For brevity here we will denote them as follows
\begin{equation}
\begin{aligned}
[N-1,N-1,1,1] &= B\,,\qquad &
[N-2,2] &= C\,,\\
[N-2,1,1] &= D\,,\qquad & [N-1,N-1,2] &= \bar D\,,\qquad & \\
[N-1,1]_\pm &= A_\pm\,, \qquad & \bullet &= S\,.
\end{aligned}
\end{equation}
For the above representations, the branching rules of the inclusion $SU(N) \subset O(N')$ read
\begin{equation}
S \to S\,,\qquad T \to A_+ \oplus B \oplus C\,,\qquad A \to A_- \oplus D \oplus \bar{D}\,.
\end{equation}
Notice that this is also in agreement with the parity of the spins exchanged.\footnote{In fact, it is the spin parity that allows us to say that $A$ goes in $A_-$ and $T$ goes in $A_+$.} Accordingly, any $O(N')$ is a special $SU(N)$ theory satisfying the extra constraints
\begin{equation}
\Delta_{A_+} = \Delta_B = \Delta_C\,,\qquad \Delta_{A_-} = \Delta_D\,,
\end{equation}
together with further constraints on OPE coefficients which we do not need for this argument. Let us call $\Delta_R^*[O(N')]$ the upper bound on the lightest operator transforming in $R$ in a $O(N')$ theory and $\Delta_R^*[SU(N)]$ the analogous quantity for $SU(N)$. Since any $O(N')$ solution is, in particular, a $SU(N)$ solution, we immediately conclude that
\begin{equation}
\Delta_R^*[SU(N)] \geq \left\lbrace\,\begin{aligned}
\Delta_S^*[O(N')]\quad & R =S\,,\\
\Delta_T^*[O(N')]\quad &R \in \{A_+,B,C\}\,,\\
\Delta_A^*[O(N')]\quad &R \in \{A_-,D\}\,.
\end{aligned}\right.\label{eq:DeltaSUNlarger}
\end{equation}
Now we would like to prove that, for the singlet, also the converse is true. To do that we need to reason in terms of the dual problem, i.e. in terms of positive functionals acting on the crossing equations. The equations for $O(N')$ can be written as
\begin{equation}
\sum_{\Delta,\ell} a_{\Delta,\ell}^{(S)}\left(\begin{matrix}
0 \\ F_{\Delta,\ell} \\ H_{\Delta,\ell}
\end{matrix}\right) + \sum_{\Delta,\ell} a_{\Delta,\ell}^{(T)}\left(
\begin{matrix}
F_{\Delta,\ell} \\ \left(1-\frac{2}{N'}\right) F_{\Delta,\ell} \\ - \left(\frac2{N'} + 1\right) H_{\Delta,\ell}
\end{matrix}
\right) 
 + \sum_{\Delta,\ell} a_{\Delta,\ell}^{(A)}\left(
\begin{matrix}
-F_{\Delta,\ell} \\ F_{\Delta,\ell} \\ -  H_{\Delta,\ell}
\end{matrix}
\right) 
=0\,.
\end{equation}
where $a_{\Delta,\ell}^{(R)}$ are the OPE coefficients squared of an operator in the representation $R$. We can combine the vectors into a matrix
\begin{equation}
M_{O(N')}  =\left(
\begin{matrix}
0 & F_{\Delta,\ell} & -F_{\Delta,\ell}
\\ F_{\Delta,\ell} &  \left(1-\frac{2}{N'}\right) F_{\Delta,\ell} & F_{\Delta,\ell} 
\\ H_{\Delta,\ell} &  - \left(\frac2{N'} + 1\right) H_{\Delta,\ell} & -  H_{\Delta,\ell}
\end{matrix}
\right)\,.
\end{equation}
In the dual formulation we look for functionals $\vec\alpha = (\alpha_1\;\alpha_2\;\alpha_3)$ such that the following vector is component-wise positive
\begin{equation}
\bigl(\alpha_1\;\;\alpha_2\;\;\alpha_3\bigr) \cdot M_{O(N')}\equiv \bigl(\alpha_S\;\;\alpha_T\;\;\alpha_A\bigr) > 0\,.
\end{equation}
If such a functional exists, then the assumptions made in the sums over $\Delta$ are inconsistent and the theory is excluded. Similarly, let $M_{SU(N)}$ be the matrix built out of the vectors given in~\eqref{eq:CrossingEquationsSing}. A positive functional $\vec\beta$ is such that the following inequalities are satisfied
\begin{equation}
\bigl(\beta_1\;\;\beta_2\;\;\beta_3\;\;\beta_4\;\;\beta_5\;\;\beta_6\bigr) \cdot M_{SU(N)}\equiv \bigl(\beta_B\;\;\beta_D\;\;\beta_C\;\;\beta_{A_+}\;\;\beta_{A_-}\;\;\beta_S\bigr) > 0\,.
\end{equation}
And, again, the existence of a positive functional implies that the $SU(N)$ theory is inconsistent. Our goal is to show that for every positive functional $\vec\alpha$ we can construct a corresponding positive functional $\vec\beta$. If we achieve this, then we obtain a set of inequalities in the opposite direction as before, namely
\begin{equation}
\begin{aligned}
\Delta_S^*[O(N')] &\geq \Delta_S^*[SU(N)]\,,\\
\Delta_T^*[O(N')] &\geq \max_{R \in \{A_+,B,C\}}\bigl(\Delta_R^*[SU(N)]\bigr)\,,\\
\Delta_A^*[O(N')] &\geq \max_{R \in \{A_-,D\}}\bigl(\Delta_R^*[SU(N)]\bigr)\,.
\end{aligned}\
\end{equation}
This, together with~\eqref{eq:DeltaSUNlarger} implies $\Delta_S^*[O(N')] = \Delta_S^*[SU(N)]$ as claimed.

We are looking for a linear relation between $\alpha$ and $\beta$. The branching rules suggest an ansatz of the form
\begin{equation}
\bigl(\beta_B\;\;\beta_D\;\;\beta_C\;\;\beta_{A_+}\;\;\beta_{A_-}\;\;\beta_S\bigr) = 
\bigl(x_1\lsp\alpha_T\;\;x_2\lsp\alpha_A\;\;x_3\lsp\alpha_T\;\;x_4\lsp\alpha_T\;\;x_5\lsp\alpha_A\;\;\alpha_S\bigr)\,. \label{eq:alphabetaAnsatz}
\end{equation}
We can solve this for $\beta_i$ requiring that the terms with $F_{\Delta,\ell}$ and $H_{\Delta,\ell}$ do not mix. There is a unique solution of the $x_i$'s that admits a linear map $\beta_i = T_i^{\;j} \alpha_j$ and it reads\footnote{For completeness we also report the solution for $T_i^{\;j}$
\[
T_i^{\;j} = \left(
\begin{array}{ccc}
 \frac{(N-1) N^2 (N+3)}{4 \left(N^2-2\right) \left(N^2+1\right)} & \frac{N^2 (N+3) \left(N^2-3\right)}{4 (N+1) \left(N^2-2\right) \left(N^2+1\right)} & 0 \\
 -\frac{(N-2) (N+2)}{N^2-2} & \frac{(N-2) (N+2)}{N^2-2} & 0 \\
 \frac{(N-3) N^2 (N+1)}{\left(N^2-2\right) \left(N^2+1\right)} & \frac{(N-3) N^2 \left(N^2-3\right)}{(N-1) \left(N^2-2\right) \left(N^2+1\right)} & 0 \\
 \frac{2 (N-2) (N-1) (N+1) (N+2)}{N \left(N^2-2\right) \left(N^2+1\right)} & \frac{2 (N-2) (N+2) \left(N^2-3\right)}{N \left(N^2-2\right) \left(N^2+1\right)} & 0 \\
 0 & 0 & -\frac{N^2 (N+3)}{4 (N+1) \left(N^2-2\right)} \\
 0 & 0 & -\frac{(N-2) (N+2)}{N^2-2} \\
\end{array}
\right)\,.
\]
\label{foot:T}}
\begin{equation}
\begin{aligned}
&x_ 1= \frac{(N-1) N^2 (N+3)}{4 \left(N^2-2\right) \left(N^2+1\right)}\,,\quad x_ 2= \frac{(N-2) (N+2)}{N^2-2}\,,\quad x_ 3= \frac{(N-3) N^2 (N+1)}{\left(N^2-2\right) \left(N^2+1\right)}\,,\\
&x_ 4= \frac{2 (N-2) (N-1) (N+1) (N+2)}{N \left(N^2-2\right) \left(N^2+1\right)}\,,\quad x_ 5= \frac{2 N}{N^2-2}\,.
\end{aligned}
\end{equation}
Note that all the $x_i$'s are nonnegative for $N \geq 3$. As a consequence of equation~\eqref{eq:alphabetaAnsatz} a positive functional $\vec\alpha$ will lead to a positive functional $\vec\beta$, which is what we wanted to show.

To conclude, let us revisit the direction of the inequality that we showed at the beginning, namely~\eqref{eq:DeltaSUNlarger}. Instead of presenting a purely group theoretic argument, one could have proceeded exactly in the same way as above, via a dual formulation of the problem. This time we are looking for a linear map from $\vec\beta$ to $\vec\alpha$. Based on the branching rules we can propose an ansatz of the following form
\begin{equation}
\bigl(\alpha_S\;\;\alpha_T\;\;\alpha_A\bigr) = 
\bigl(\beta_S\quad y_1\lsp\beta_{A_+} \! + y_2 \lsp\beta_{B}+y_3\lsp\beta_C \quad  y_4\lsp\beta_{A_-}\!+y_5\lsp\beta_D\bigr)\,. 
\end{equation}
As before, we find a solution with nonnegative $y_i$'s
\begin{equation}
y_1 = \frac{N}{N^2-4}\,,\qquad y_2 = 2\,,\qquad y_3 = \frac12\,,\qquad y_4 = \frac1{N}\,,\qquad y_5 = 1\,.
\end{equation}
Let $\tilde{T}$ be the resulting linear transformation $\alpha_j = \tilde{T}_j^{\;i}\beta_i$. It is easy to check that it is just the left inverse of the transformation shown in footnote~\ref{foot:T}, namely $\tilde T \cdot T = \mathds1_3$.

\section{Parameters of the numerical implementation}
\newcommand{\as}[1]{\renewcommand{\arraystretch}{#1}}

The numerical conformal bootstrap problem was truncated according to the parameters in Table~\ref{table:truncationBootstrap}. The semi-definite problem was solved using sdpb \cite{Simmons-Duffin:2015qma,Landry:2019qug} with the choice of parameters given in Table~\ref{table:sdpb}.

\begin{table*}[h!]
	\centering
	\as{1.2}
	\begin{tabular}{@{}l l l@{}}
		\toprule
		& $\Lambda=19$  & $\Lambda=27$ \\
		\midrule
		\texttt{Lset} & $\{0,...,26\}\cup\{49,50\}$ & $\{0,...,30\}\cup\{39,40,49,50\}$ \\
		\texttt{order}  & 60 & 60 \\
		$\kappa$ & 14 & 18\\
		\bottomrule 
	\end{tabular}
	\caption{Values of the various parameters appearing in the numerical bootstrap problem.}
	\label{table:truncationBootstrap}
\end{table*}

\begin{table}[H]
	\centering
	\as{1.2}
	\begin{tabular}{@{}l l l@{}}
		\toprule
		Parameter & feasibility & OPE \\
		\midrule
		\texttt{maxIterations} & $500$ & $500$ \\
		\texttt{maxRuntime} &  $86400$&  $86400$ \\
		\texttt{checkpointInterval} &  $3600$&  $3600$ \\
		\texttt{noFinalCheckpoint} & \texttt{True} &  \texttt{False} \\
		\texttt{findDualFeasible} & \texttt{True} &  \texttt{False} \\
		\texttt{findPrimalFeasible} & \texttt{True} &  \texttt{False} \\
		\texttt{detectDualFeasibleJump} & \texttt{True} &  \texttt{False} \\
		\texttt{precision} &  $700$&  $700$ \\
		\texttt{maxThreads} &  $28$&  $28$ \\
		\texttt{dualityGapThreshold} &  $10^{-20}$&  $10^{-20}$ \\
		\texttt{primalErrorThreshold} &  $10^{-60}$&  $10^{-60}$ \\
		\texttt{dualErrorThreshold} &  $10^{-60}$&  $10^{-60}$ \\
		\texttt{initialMatrixScalePrimal} &  $10^{20}$&  $10^{20}$ \\
		\texttt{initialMatrixScaleDual} &  $10^{20}$&  $10^{20}$ \\
		\texttt{feasibleCenteringParameter} &  $0.1$&  $0.1$ \\
		\texttt{infeasibleCenteringParameter} &  $0.3$&  $0.3$ \\
		\texttt{stepLengthReduction} &  $0.7$&  $0.7$ \\
		\texttt{choleskyStabilizeThreshold} &  $10^{-40}$&  $10^{-40}$ \\
		\texttt{maxComplementarity} &  $10^{200}$&  $10^{200}$ \\
		\bottomrule
	\end{tabular}
	\caption{Parameters used in \texttt{sdpb} for respectively feasibility problems and for OPE optimization.}\label{table:sdpb}
\end{table}

\bibliographystyle{utphys}
\clearpage
\bibliography{references}

\end{document}